\documentclass[acmsmall]{acmart}

\AtBeginDocument{%
 \providecommand\BibTeX{{%
 \normalfont B\kern-0.5em{\scshape i\kern-0.25em b}\kern-0.8em\TeX}}}

\copyrightyear{2020}
\acmYear{2020}
\usepackage{eurosym}
\acmJournal{CSUR}
\usepackage{pifont}
\newcommand{\cmark}{\ding{51}}%
\newcommand{\xmark}{\ding{55}}%
\renewcommand{\textrightarrow}{$\rightarrow$}%
\usepackage{url}
\usepackage[linesnumbered,ruled,vlined,noend]{algorithm2e}
\usepackage{xcolor}
\usepackage{colortbl}
\usepackage{tabularx}
\usepackage{threeparttable}
\usepackage{booktabs}
\usepackage{cellspace}
\usepackage{makecell}
\usepackage{multirow}
\usepackage{natbib}
\usepackage{url}
\usepackage[group-separator={,}]{siunitx}
\usepackage{rotating}
\usepackage{wrapfig}
\usepackage{graphicx,subfigure}
\usepackage{tikz}
\usetikzlibrary{tikzmark}
\usetikzlibrary{decorations.pathreplacing,shapes,snakes}
\usetikzlibrary{arrows}

\usepackage{wrapfig}
\usepackage{floatflt}

\usepackage{enumitem}
\renewcommand{\cite}{\citep}

\newcommand{\shane}[1]{\textrm{\textcolor{orange}{Shane says: #1}}}
\newcommand{\myparagraph}[1]{\vspace{0.6\baselineskip}\noindent{\textbf{#1.}}~}
\newcommand{\var}[1]{\mbox{\emph{#1}}}
\newcommand{\svar}[1]{\mbox{\scriptsize\emph{#1}}}
\newcommand{\tvar}[1]{\mbox{\tiny\emph{#1}}}
\newcommand{\avar}[1]{\mbox{#1}}

\newtheorem{definition}{Definition}

\newcommand{\numb}[1]{\num[group-separator={,}]{#1}}

\newcommand{\pats}{{\footnotesize\textsf{PATS}}\xspace}

\newcommand{\tkstt}{{\small\textbf{T$k$STT}}\xspace}

\DeclareMathOperator*{\argmax}{arg\,max}
\DeclareMathOperator*{\argmin}{arg\,min}
\newcommand{\traclus}{{\small{\textsf{TRACLUS}}}\xspace}
\newcommand{\ebd}{{\footnotesize{\textsf{EBD}}}\xspace}

\newcommand{\kpath}{{\small{\textbf{$k$-paths}}}\xspace}
\newcommand{\kmeans}{{\small{\textbf{$k$-means}}}\xspace}

\newcommand{\inverted}{II}

\SetKwComment{Comment}{$\triangleright$\ }{}

\newcommand{\query}{{\small\textbf{MASK}}\xspace}
\newcommand{\tksk}{{\small\textbf{T$k$SK}}\xspace}

\newcommand{\kbct}{{\small\textbf{$k$BCT}}\xspace}

\newcommand{\rknnt}{{\small\textbf{R${k}$NNT}}}
\newcommand{\knn}{{\small\textbf{${k}$NN}}\xspace}

\newcommand{\ra}[1]{\renewcommand{\arraystretch}{#1}}

\newcommand{\rknn}{{\small\textbf{R${k}$NN}}\xspace}
\newcommand{\tran}{{T}}

\newcommand{\destination}{v_e}

\newcommand{\maxrknnt}{{\small\textbf{MaxR${k}$NNT}}}

\newcommand{\scomment}[1]{\iffalse#1\fi}

\SetKwComment{Comment}{$\triangleright$\ }{}

\newcommand{\lors}{{\footnotesize{\textsf{LORS}}}\xspace}
\newcommand{\lcss}{{\footnotesize{\textsf{LCSS}}}\xspace}
\newcommand{\dtw}{{\footnotesize{\textsf{DTW}}}\xspace}
\newcommand{\erp}{{\footnotesize{\textsf{ERP}}}\xspace}
\newcommand{\edr}{{\footnotesize{\textsf{EDR}}}\xspace}

\usepackage{marginnote}
\makeatletter
\let\oldmarginnote\marginnote
\renewcommand*{\marginnote}[1]{%
	\begingroup%
	\ifodd\value{page}
	\if@firstcolumn\normalmarginpar\fi
	\else
	\if@firstcolumn\else\normalmarginpar\fi
	\fi
	\oldmarginnote{\textcolor{red}{#1}}%
	\endgroup%
}
\makeatother
\usepackage{xcolor}
\usepackage{color,soul}
\setul{0.5ex}{0.3ex}
\definecolor{Green}{rgb}{0,1,0}
\setulcolor{Green}

\begin{document}
\sloppy
\title{A Survey on Trajectory Data Management, Analytics, and Learning}


\author{Sheng Wang}
\email{swang@nyu.edu}
\orcid{0000-0002-5461-4281}
\affiliation{%
 \institution{New York University}
 \city{}
 \country{United States}
}

\author{Zhifeng Bao}
\email{zhifeng.bao@rmit.edu.au}
\author{J. Shane Culpepper}
\email{shane.culpepper@rmit.edu.au}
\orcid{0000-0002-5461-4281}
\affiliation{%
	\institution{RMIT University}
	\country{Australia}
}

\author{Gao Cong}
\email{gaocong@ntu.edu.sg}
\orcid{0000-0002-1902-9087}
\affiliation{%
	\institution{Nanyang Technological University}
	\country{Singapore}
}
\renewcommand{\shortauthors}{Wang, et al.}

\begin{abstract}
{
Recent advances in sensor and mobile devices have enabled an
unprecedented increase in the availability and collection of urban
trajectory data, thus increasing the demand for more efficient ways
to manage and analyze the data being produced.
In this survey, we comprehensively review recent research trends in
trajectory data management, ranging from trajectory pre-processing,
storage, common trajectory analytic tools, such as querying spatial-only
and spatial-textual trajectory data, and trajectory clustering.
We also explore four closely related analytical tasks commonly used
with trajectory data in interactive or real-time processing.
Deep trajectory learning is also reviewed for the first time.
Finally, we outline the essential qualities that a trajectory data
management system should possess in order to maximize flexibility.
}
\end{abstract}



\maketitle

\section{Introduction}
\begin{figure}
	\subfigure[{\scriptsize {Human trajectories in surveillance} \cite{Zhou2012}} ]{\includegraphics[height=2.7cm]{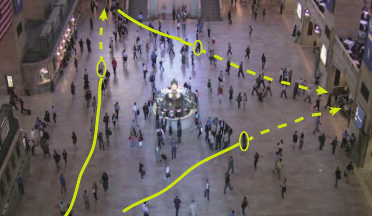}}~~~~~~~~~~~~~~~~~~~~
	\subfigure[ \scriptsize Monitoring cars by UAV \cite{Bock2019}]{\includegraphics[height=2.7cm]{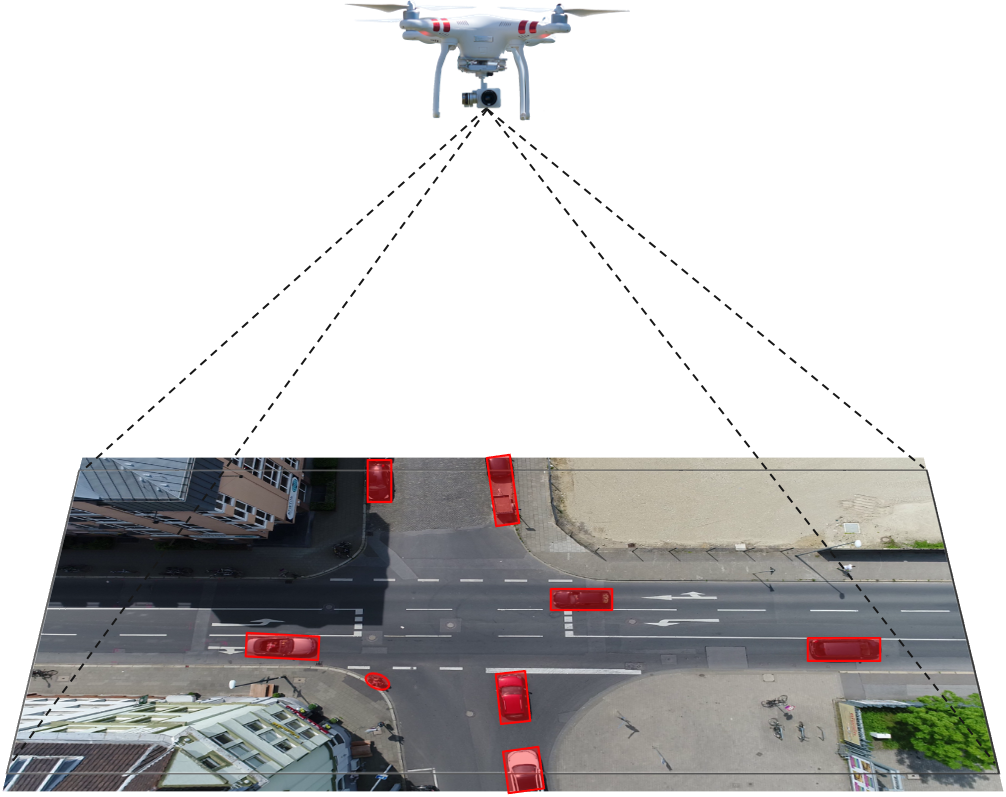}}~~~~~~~~~~~~~~~~~~~~
	\subfigure[\scriptsize  Car localization by RFID \cite{rfid}]{\includegraphics[height=2.7cm]{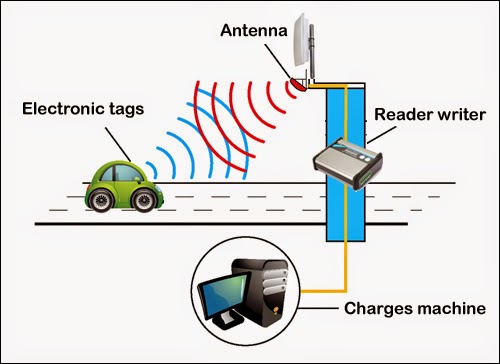}}
	\vspace{-1.5em}
	\caption{New localization techniques that can capture trajectory data more proactively than using GPS.}
	\vspace{-1.5em}
	\label{fig:more-devices}
\end{figure}

Just over twenty years ago, Global Positional System (GPS) satellites
began transmitting two additional signals to be used for civilian
(non-military) applications to improve aircraft
safety.\footnote{\url{https://www.pcworld.com/article/2000276/a-brief-history-of-gps.html}}
Demand for GPS-based navigation has grown steadily ever since,
and privately owned vehicles are now the dominant consumers of this
technology.
An increased reliance on GPS-equipped smartphones has also led to a
rise in the use of location-based services (\textbf{LBS}), such as
ride-sharing and social-network check-ins.

In addition to GPS data, other sensor devices such as traffic
surveillance cameras, unmanned aerial vehicles
(UAV),\footnote{\url{https://www.gpsworld.com/in-mit-project-drones-map-without-gps/}}
and Radio-frequency identification (RFID) can also collect location
data without deploying battery-dependent receiving devices to track
objects, as shown in Fig.~\ref{fig:more-devices}.
Such devices can accurately report precise locations and track
objects proactively and continuously.

\subsection{Trajectory Data}
Devices that track moving objects often generate a series of
ordered points representing a \textit{trajectory}.
More formally, a trajectory $T$ is composed of two or more
\textit{spatial-only} points and defined as:

\begin{definition}{\textbf{(Point)}}
	\label{def:point} 
A point $p = \{\avar{x}, \avar{y}, t\}$ records the latitude
$\avar{x}$, the longitude $\avar{y}$ at timestamp $t$.
\end{definition}

\begin{definition}{\textbf{(Trajectory)}}
{A trajectory $T$ is a sequence of points
$\{(p_1, \gamma_1), (p_2, \gamma_2),\ldots,(p_{m}, \gamma_m)\}$}.
\end{definition}


The number of points in a trajectory $T$ is denoted as the
\textit{length} of the trajectory, and the \textit{sampling rate} is
the number of samples per second (or other time units).
Additional information $\gamma$ can be included with each point $p$ in a
trajectory $T$ generated from the location-based service.\footnote{{For spatial-only trajectories, we can set $\gamma=\emptyset$. Note that $\gamma$ can also be attached to the whole trajectory directly.}}
For example, textual data can be integrated into a trajectory from
social network check-in data or travel blogs, and has been referred
to as
\textit{spatial-textual trajectories} \cite{Kim2015b,Bao2012Location}, \textit{semantic trajectories} \cite{Zhang2014a}, or \textit{symbolic trajectories}~\cite{Damiani2015}.
This survey will focus primarily on these two closely related forms
of trajectory data as solutions and applications in recent research
work commonly overlap.

\myparagraph{Survey Scope} 
We focus primarily on urban trajectories.
Other studies of domain specific data such as aircraft
\cite{Ayhan2016} are not covered in this work. 
Note that the main difference between a moving object
\cite{Saltenis2000b} and a trajectory is the distinction between
a pointwise sequence and a continuous projection, and each of them
often employs fundamentally different storage and algorithmic
solutions.

\myparagraph{Public Trajectory Datasets}
Table~\ref{tab:alldatasets} shows several existing urban trajectory
datasets, which can be broadly divided into three groups: humans,
vehicles (car, truck, train, bus, tram, etc.), others (animals and
hurricanes).
We can observe that human derived 
trajectory data (including vehicles) is a common source of
trajectory data,
and is currently the largest source of trajectory data.
For example, $1.1$ billion (seventh row) individual taxi trips were
recorded in NYC from January 2009 through June 2015.
Other datasets including the movement of animals and hurricanes have
relatively few available trajectories, making such collections
much more tractable to process and analyze.


\begin{table}
	\ra{1}
	\centering
	\caption{Publicly available trajectory datasets.}
	\vspace{-1em}
	\label{tab:alldatasets}
	\scalebox{0.75}{\begin{tabular}{cccccc}
			\cmidrule{1-5}\addlinespace[-2.7pt]\rowcolor{gray!30}
			\textbf{Categorization} &\textbf{Type} & \textbf{Exemplary Datasets}& \textbf{\#Trajectories} & \textbf{Applications} \\ \addlinespace[0pt]\cmidrule{1-5}
			\multirow{5}{*}{Human}	& {GPS tracking} & GeoLife \cite{Zheng2009} & \numb{17621} & Human mobility \cite{Zhang2016a} &
			\multirow{12}{*}{\hspace{-2em}$\left.\begin{array}{l}
				\\
				\\
				\\ \\ \\ \\ \\ \\ \\ \\ \\ \\ 
				\end{array}\right\rbrace{}\makecell{\text{Urban}\\\text{area}} $} 
			\\ 
			& Check-in & \makecell{Foursquare \cite{Bao2012Location}\\ Gowalla \cite{Cho2011}} & \numb{104478} & \makecell{Tourism planning \cite{Wang2017} \\Mobility prediction \cite{Jin2018}}\\ 
			&Online sharing & \makecell{OpenStreetMap \cite{osm} }& 8.7 million & Commuting analytics \\
			&{Video surveillance} & {Grand central station \cite{Zhou2012}}& \numb{20000} &Crowd behavior \\
			&{COVID-19} & \makecell{KCDC \cite{covid19-2}, JHU \cite{covid19}}& <100 & Disease control \\
			 
		\cmidrule{1-5}
			\multirow{4}{*}{Vehicle}&Taxi trips & T-drive \cite{Yuan2010}, {Porto} \cite{portodata} & 1.7 million & Traffic monitoring\\ 
			&Taxi trip-requests & NYC \cite{nyctaxidata}, Chicago \cite{nyctaxidata}& 1.1 billion& Ride sharing\\ 
			&{Traffic cameras} & NGSIM \cite{NGSIM} & - & Traffic simulation\\
			&{Drones} & HighD \cite{Krajewski2018}, inD \cite{Bock2019}& \numb{110000} & Traffic prediction\\
			& {Self-driving}& Argoverse \cite{Chang2019}, ApolloScape \cite{Ma2018} & \numb{300000} & Trajectory forecasting \\
			&Trucks & Greece Trucks \cite{greecetrucks} & \numb{1100} & Pattern mining\\ \cmidrule{1-5}
			\multirow{2}{*}{Others}&Hurricanes & Atlantic hurricanes~\cite{nhcdata}& \numb{1740} &Disaster detection \\ 
			&Animals & Zebranet \cite{nhcdata}, Movebank \cite{Movebank}& 33 & Animal behavior \cite{LaPoint2013}\\
			\cmidrule{1-5} 
	\end{tabular}}
	\vspace{-1.4em}
\end{table}

\subsection{An Overview}
\label{sec:overview}
\begin{figure}
	\centering
	\includegraphics[width=14cm]{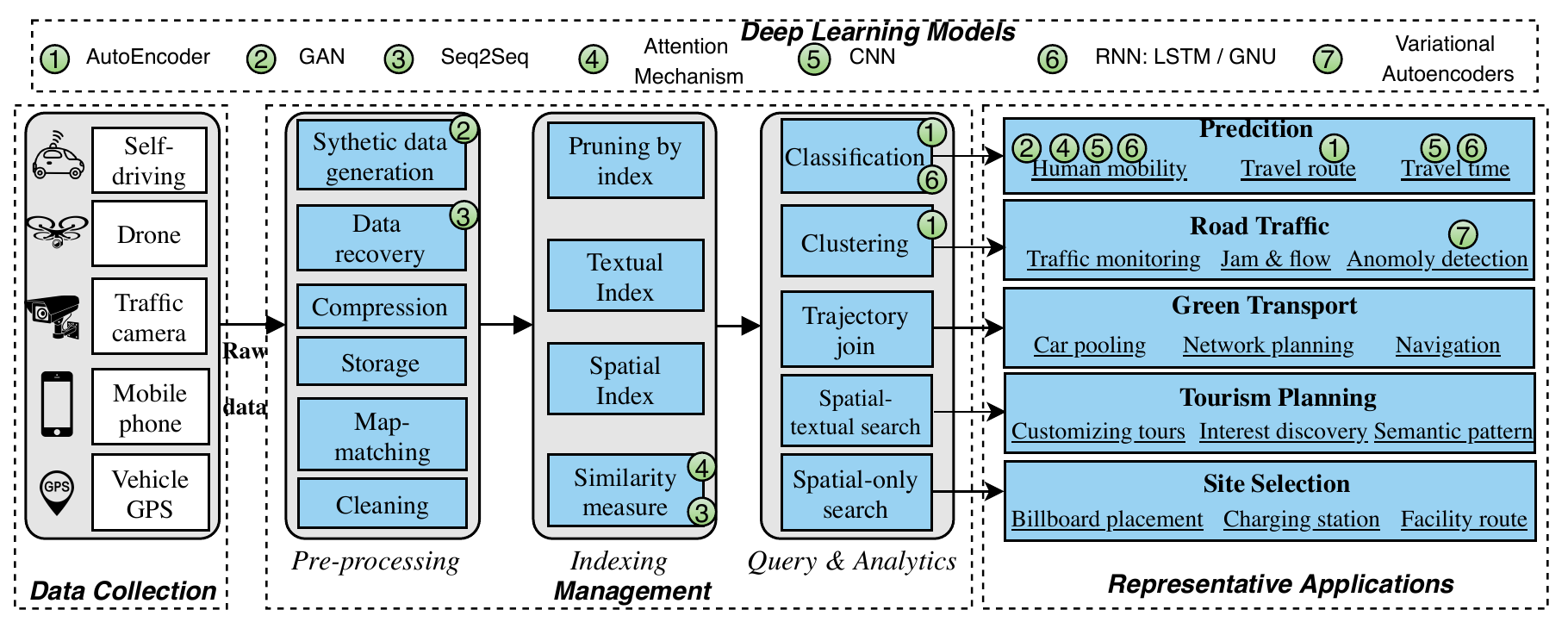}
	\vspace{-2.5em}
	\caption{{Overview of trajectory data collection, management,
		representative applications, and widely used deep learning models (indicated by the green labels, and the full names of models can be found in Section~\ref{sec:deep})}.}
	\vspace{-1em}
	\label{fig:data-query}
\end{figure}

Fig.~\ref{fig:data-query} provides an overview of trajectory data
management, analytics, and learning. 
Briefly, a trajectory data management and analytic system
has several fundamental components:

\begin{itemize}
\item \textbf{Cleaning}: Common techniques to clean common sensor
errors in raw trajectories and normalize sampling rates.

\item \textbf{Storage}: Open-source, commercial, and other commonly
deployed storage formats that are used to represent trajectory data.

\item \textbf{Similarity Measures}: 
The most commonly used measures for top-$k$ similarity search,
similarity joins, and clustering.

\item \textbf{Indexing}: State-of-the-art solutions for indexing
spatial and spatial-textual trajectory data, as well as effective
pruning solutions to improve performance.

\item \textbf{Query and Analytics}: The most common trajectory queries, including
range queries, top-$k$ similarity search, and trajectory joins.


\item \textbf{Upstream Urban Applications}: The four most common offline urban
applications of trajectory data with problem formulations and
solutions.
\end{itemize}

\begin{table}
	\centering
	\caption{Survey contributions relative to prior trajectory surveys.}
	\vspace{-1em}
	\label{tab:distinction}
	\scalebox{0.75}{
	\begin{tabular}{lccccccc}
		\toprule
		\addlinespace[0pt]\rowcolor{gray!30}
		\textbf{Survey} & \textbf{Datasets} & \textbf{\makecell{Commercial\\ systems}} & \textbf{\makecell{Similarity\\measures}} & \textbf{\makecell{Pruning\\mechanisms}} & \textbf{Indexing} & \textbf{\makecell{Processing\\ pipeline \\decomposition}} & \textbf{\makecell{Deep\\ learning} } \\
		\addlinespace[0pt]\midrule
		2011 \cite{Zheng2013a} & \xmark & \xmark & spatial & spatial & spatial & \xmark & \xmark \\
		2013 \cite{Parent2013a} & \xmark & \xmark & \xmark & \xmark & \xmark & \xmark & \xmark \\
		2015 \cite{Zheng2015b} & only GPS & \xmark & spatial & \xmark & spatial & mining & \xmark \\
		2016 \cite{Feng2016a} & \xmark & \xmark & \xmark & \xmark & \xmark & \xmark & \xmark \\
		2018 \cite{Richly2018} & \xmark & \xmark & \xmark & \xmark & spatial & \xmark & \xmark \\ 
		\midrule
\textbf{\makecell{This survey}} & \makecell{ \cmark } & \makecell{\cmark } & \makecell{\cmark } & \makecell{\cmark} & \makecell{\cmark } & \makecell{ \cmark} & \makecell{\cmark } \\ \bottomrule
	\end{tabular}}
	\vspace{-1.5em}
\end{table}

\subsection{New Expository Contributions Found from this Survey}
\label{sec:new}
Managing large-scale trajectory data has many important challenges
in efficiency and scalability due to its size and diversity.
A survey covering state-of-the-art and open problems in each stage, from data pre-processing all the way to upper-stream applications, is valuable to both research and industrial communities.
While several surveys on trajectory data exist, none review
trajectory data management and applications as outlined in
Fig.~\ref{fig:data-query}.
Prior surveys have covered trajectory data mining \cite{Zheng2015b},
semantic trajectory modeling \cite{Parent2013a}, and related
applications \cite{Richly2018}.
Table~\ref{tab:distinction} compares and contrasts prior surveys with
our own, highlighting the new areas covered.
A couple of surveys from closely related areas which do partially
overlap with the content of this survey include: trajectory
similarity measures \cite{Su2019b}, trajectory privacy
\cite{Guo2015b}, and trajectory classification
\cite{Bian2019}.
In these instances, this survey includes more recent work from these
important areas.
The closest related review work is the textbook ``Computing with spatial
trajectories'' \cite{Zheng2013a}, published in 2011.
Current techniques have advanced significantly in the last ten
years, including the quickly growing area of spatial-textual trajectories,
which were not covered previously.
The key contributions of this survey can be summarized as:

\begin{itemize}

\item \textbf{Scalability and Storage}.
This survey comprehensively reviews trajectory storage, which is
commonly required in scalable offline analytics.

\item \textbf{Processing Pipeline Decomposition}. 
We provide insights into the connections among all components in
the entire data processing pipeline.
For the backend, we compare trajectory storage systems (academic and
commercial); for middleware, we compare similarity measures, queries,
and index structures; for upstream trajectory applications, we
propose a taxonomy for key operators used across multiple
tasks.

\item \textbf{Representative Applications}.
Based on a comprehensive review of recent research in SIGMOD, PVLDB,
ICDE, KDD, SIGSPATIAL, VLDBJ, TKDE, IJGIS, and TITS, we focus on
urban applications requiring real-time responses or timely decision
making, including: 1) road traffic; 2) green transport; 3) tourism
planning; 4) site selection.
These applications depend heavily on downstream analytics tools.
For example, clustering is commonly used to generate candidate routes
and to generate a fixed number of results for traffic monitoring.

\item \textbf{Deep Trajectory Learning}.
New trends on using deep learning for various trajectory data
analytics tasks, including trajectory recovery and generation,
trajectory representation and similarity search, clustering,
classification, anomaly detection, and trajectory prediction.

\item \textbf{Perspectives on Future Trajectory Data Management Challenges}.
We conclude with a list of key challenges and associated open problems
for researchers working in the area.
\end{itemize}

\subsection{Outline of This Survey}
The remainder of the survey is organized as follows:
Section~\ref{sec:manage} presents techniques for trajectory data
storage and pre-processing.
Section~\ref{sec:distance} compares common trajectory similarity
measures.
Section~\ref{sec:spatialonly} illustrates
queries and indexing techniques for spatial-only trajectory data and
spatial-textual trajectory data, respectively.
Section~\ref{sec:clustering} reviews trajectory clustering from a
data management perspective.
Section~\ref{sec:application} presents a taxonomy of four application
types commonly encountered in trajectory data management.
Section~\ref{sec:deep} introduces the latest progress in deep
trajectory learning.
Section~\ref{sec:future} discusses the future of trajectory data
system management.
Section~\ref{sec:conclude} concludes the paper.

\section{Management Systems and Pre-processing}
\label{sec:manage}

\subsection{Trajectory Representation and Storage}
A list of coordinates denoted by two floats (e.g.,
{\small``-37.807302, 144.963242''}) are the traditional representation
of a spatial point location.
Two or more such points can be used to represent a trajectory for
storage, search, and analytics, and is the most common method in both
research and industry.
Table~\ref{tab:systems} shows several representative research systems
specifically designed for trajectory data, as well as several
commercial systems which were not necessarily designed for only
trajectory data, but can store and manipulate trajectory data through
module extensions.
Since commercial systems are designed to handle many different types
of data, they can incur additional cost overheads.

In the academic field, several trajectory storage systems have been
built to manage trajectory data
\cite{Cudre-Mauroux2010,Wang2014e,Shang2018,Ding2018,Xie2016a,Wangsheng2018}, and
store trajectories as a set of points.
These trajectory management systems
\cite{Cudre-Mauroux2010,Wang2014e,Xie2016a} only support a single
storage format, and often only a single query type such as a range
query (finding trajectories in a rectangle).
More recently, distributed systems \cite{Shang2018,Li2020,Ding2018} have
been developed to store point-based trajectory datasets scalably, and
can perform more advanced queries such as $k$NN search over
trajectories.

Existing commercial or open-source systems (e.g., a general database
such as Oracle
Spatial\footnote{\url{https://www.oracle.com/technetwork/database/options/spatialandgraph/overview/index.html}},
SQL
Server\footnote{\url{https://docs.microsoft.com/en-us/sql/relational-databases/spatial/spatial-data-sql-server}},
and
MySQL\footnote{\url{https://dev.mysql.com/doc/refman/5.7/en/gis-linestring-property-functions.html}},
or geospatial databases such as
ArcGIS\footnote{\url{http://www.arcgis.com/index.html}},
Tile38\footnote{\url{https://github.com/tidwall/tile38}},
PostGIS\footnote{\url{https://postgis.net/}},
GeoMesa\footnote{\url{https://www.geomesa.org/}}) can
also be extended for trajectory management.
For example, {\em LineString} defined
a portable data-interchange
format-GeoJSON\footnote{\url{https://en.wikipedia.org/wiki/GeoJSON}}
which is a geographical representation for an array of point coordinates.
A trajectory can be stored as a {\em LineString} using this format.
\textit{GPX} (the GPS Exchange Format) \cite{foster2004gpx} is
another format used by OpenStreetMap for routing and tracking.
Figures~10 and ~11 in Appendix~\cite{Wang2020} show an example of
these two formats.
However, operator support for LineString and GPX data is limited,
e.g., return the number of points, the starting the ending point, or
the $i$-th point.
Range and \knn queries are supported over all points, but not the
trajectories themselves.
%
%
\begin{table}
	\centering
	\ra{1}
	\caption{An overview of existing trajectory database systems.}
	\vspace{-1em}
	\scalebox{0.75}{\begin{tabular}{cccccc}
		\toprule
		\addlinespace[0pt]\rowcolor{gray!30}
		\textbf{System} & \textbf{Data} & \textbf{Indexing} & \textbf{Range} & \textbf{$k$NN} & \textbf{Similarity Measure} \\
		\addlinespace[-1pt]\midrule
		TrajStore \cite{Cudre-Mauroux2010} & Trajectory & Quad-tree & \cmark & \xmark & \xmark \\
		SharkDB \cite{Wang2014e} & Trajectory & Frame structure & \cmark & \cmark & Point-to-trajectory \\
		UITraMan \cite{Ding2018} & Trajectory & R-tree & \cmark & \cmark & Point-to-trajectory \\
		TrajMesa \cite{Li2020} & Trajectory & Z-order curve & \cmark & \cmark & Point-to-trajectory \\
		DITA \cite{Shang2018} & Trajectory & R-tree & \cmark & \cmark & Trajectory-to-trajectory \\ 
		Torch \cite{Wangsheng2018} & Trajectory & Grid-index & \cmark & \cmark & Trajectory-to-trajectory \\ 
		\midrule
		Oracle & LineString & R-tree & \cmark & \cmark & Point-to-point \\
		PostGIS & LineString & R-tree & \cmark & \cmark & Point-to-point \\
		SQL Server & LineString & Grid-index & \cmark & \cmark & Point-to-point \\
		GeoMesa & LineString & Z-order curve & \cmark & \cmark & Point-to-point \\
		Tile38 & LineString & R-tree & \cmark & \cmark & Point-to-point \\
		GeoSpark \cite{Yu2015} & LineString & R-tree & \cmark & \cmark & Point-to-point \\
		SpatialSpark \cite{You2015a} & LineString & R-tree & \cmark & \xmark & \xmark \\ \bottomrule
	\end{tabular}}
	\label{tab:systems}
		\vspace{-1.5em}
\end{table}

SpatialSpark \cite{You2015a} and GeoSpark \cite{Yu2015} are two
representative open-source systems from the research field of spatial
databases, which also support the storage and querying of LineString.
Range and $k$NN queries are supported for the point data, but not
trajectories.
Nevertheless, these spatial database systems could be extended
to support any number of trajectory search operators
in the future since trajectory data is an extension point data.
In Section~\ref{sec:spatialindexing}, we will review indexing
techniques for point data in more detail.

\subsection{Trajectory Pre-processing}
\label{sec:preprocess}
\myparagraph{Trajectory Cleaning}
Since trajectory data is often collected with GPS devices
that have an average user range error (URE) of 7.8 m (25.6 ft), with
95\% probability, trajectory point data is \textit{noisy}
\cite{Zheng2015b}.
Moreover, the sampling rate of GPS devices varies by application.
For example, two vehicles following an identical path can produce
trajectories that contain a different number of sampled points.
These factors can directly affect distance or
similarity computations, and degrade result quality.
Hence, when processing raw trajectory datasets, data cleaning is not
only beneficial, but in certain circumstances a requirement.
\textit{Data segmentation} \cite{Buchin2011},
\textit{calibration} \cite{Su2013} and \textit{enrichment}
\cite{Alvares2007} are the three most common cleaning techniques
for trajectory data.
Specifically, segmenting data splits a long trajectory into several
short trajectories.
For example, analysis of a single taxi ``trip'' makes more sense than
analyzing the movements of the taxi for an entire day.
\citet{Buchin2011} addressed the problem of segmenting a trajectory
based on spatiotemporal criterias, including location, heading, speed, velocity, curvature, sinuosity, curviness, and shape. 

\citet{Su2013} proposed a calibration method that transforms a heterogeneous trajectory dataset to one with (almost) unified sampling strategies, such that the similarity between trajectories can be computed more accurately.
For noisy points, calibration identifies outliers
and adds statistical correction instead of filtering out
these points directly. 
Such calibration also plays an important role in precise similarity computation \cite{ranu2015indexing}, which we will introduce later.
\citet{Liu2012a} further considered the constraint of road network topology, geometry
information, and historical information to re-calibrate noisy data
points.
For trajectories of sparse data points, \citet{Alvares2007}
proposed that the data can be enriched with additional points and
semantic information about the types of visited places.
Conversely, trajectory simplification \cite{Zhang2016e} can be
applied to remove redundant points.

\myparagraph{Trajectory Compression}
Existing trajectory compression techniques can be divided into two
groups: simplification-based and road network-based.
Excluding extra points is a common space reduction method when the
sampling rate is high, and is commonly referred to as {\em trajectory
simplification} \cite{Zhang2016e}, and also used in
trajectory cleaning as discussed above.
Removing points can reduce size, but must be used carefully as it can
also reduce the resolution when analyzing the data.
To alleviate this problem, an error ratio can be applied to bound the
loss for specific computations and data processing operations that
are predefined before data simplification~\cite{Long2013}.

Alternatively, a road network can be used to enable better
compression \cite{Song2014} with little or no
reduction in quality, for certain types of data.
Each trajectory is projected onto a road network as a sequence of
road segments.
Next, each road segment is uniquely encoded using \textit{Huffman coding}
\cite{cormen2009introduction}.
Each trajectory is succinctly represented as a concatenation
of the codewords, and is significantly more effective than
attempting to compress the raw floating point value pairs (latitude
and longitude).
Further, string compression techniques \cite{Yang2017b} can
be used directly on the trajectory data, and any temporal information
can also be succinctly stored using these techniques.

\myparagraph{Map-matching}
Using a road network, map-matching \cite{Lou2009,Newson2009} projects
a raw trajectory onto a real path, and supports cleaning and
compression.
A road network is modeled as a weighted graph
\cite{cormen2009introduction}, where a road segment is an edge, and
its weight represents the length of this road.
Consequently, a path mapped from a raw trajectory is a set of connected edges
in a road network.

\vspace{-0.5em}
\begin{definition}\textbf{(Road Network)}
A road network is a directed graph $G=(V, E)$, where $V$ is a set of
vertices $v$ representing the intersections and terminal points of
the road segments, $E$ is a set of edges $e$ representing road
segments, and each vertex has a unique id allocated from $1$ to $|V|$.
\end{definition}

\begin{definition}\textbf{(Mapped Trajectory)}
Given a raw trajectory $T$ and a road network $G$, we map ${T}$ to a
road network path $P$ which is composed of a set of connected edges
in $G$, such that ${T}:e_1\rightarrow e_2\rightarrow
\ldots\rightarrow e_{m}$, and denoted as $T$ containing a series of
edge IDs.
\end{definition}

Map-matching techniques should consider both effectiveness and
efficiency.
Efficient algorithms enable us to find the nearest road segments for
each point in the trajectory, and are required in order to connect the
candidate road segment to the best path.
Uncertainty is often the biggest hurdle, and is a reflection of 
parameter selection, which would reduce the number of potential candidate
paths during shortest path search in $G$.
For effectiveness, mapping a trajectory to the true traversal
path of a vehicle when errors in the data can be substantial is
a key limiting factor.

To measure effectiveness, a ground truth is required, and is a
problem in its own right.
There are three common ways to generate a ground truth for the
map-matching problem: 1) using real vehicles equipped with GPS
devices on the road~\cite{Newson2009}; 2) human
adjudication~\cite{Lou2009}; and 3) simulated GPS sampling
\cite{Jagadeesh2004} which is the least costly approach. 

\section{Trajectory Similarity Measures}\label{sec:distance}
This section will introduce the most widely used
similarity measures to compare two trajectories.
Fig.~\ref{fig:distance-example} presents two different types
of query-to-trajectory similarity measures.
Labels on each dashed line denote the distance from a query point to
a point in the trajectory.

\subsection{Point-to-Trajectory}
A $k$ Best Connected Trajectories search
(\kbct)~\cite{Chen2010a,Qi2015,Tang2011} is arguably the most commonly
employed formulation of point-to-trajectory similarity search.
To compute the result, each query point $q \in Q$ is paired with the
closest point $p \in T$, and the sum of the distances for all pairs
are aggregated to generate the final distance (similarity) score: $d_{\svar{kBCT}}(Q,T) = \sum_{q\in Q}{\min_{p\in T} d(q, p)}$.

\begin{figure}
	\centering
	\includegraphics[height=1.6cm]{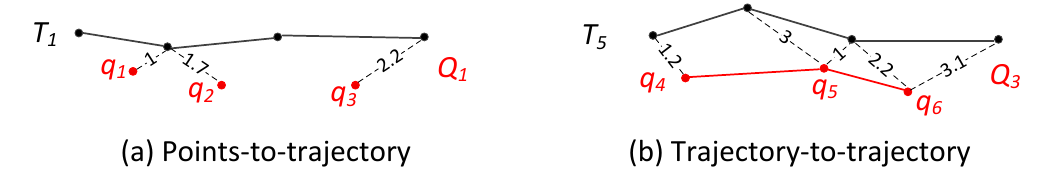}
	\vspace{-1em}
	\caption{Two types of similarity measures: (a) a query is a set of points; (b) a query is a trajectory.}
	\vspace{-1.5em}
	\label{fig:distance-example}
\end{figure}

For example, applying $d_{\svar{kBCT}}(Q,T)$ to
Fig.~\ref{fig:distance-example} yields
$d_{\svar{kBCT}}(Q_1, T_1)=1+1.7+2.2=4.9$, computed as
(a) is the sum of the distance between every query point in $Q$ to
its nearest neighbor in $T$ (the three dotted lines).
When there is only a single query point, the problem reduces to a
\knn query~\cite{Ding2018,Wang2014e}.
Since this solution requires every point in every trajectory in a
collection to be considered in order to locate the the nearest point
for all $q\in Q$, the complexity of a brute force solution is
quadratic.
Note that this distance measure can only be computed when the query
$Q$ is a set of points, and does not obey the symmetry rule, i.e.,
$d(Q,T)\ne d(T,Q)$.

An alternative to summing the distances between all nearest neighbor
pairs is another well-known measure called the \textit{closest-pair
distance (CPD)}
\cite{Zheng2015b}, which minimizes distance as follows: 
$d_{\svar{CPD}}(Q,T) = \min_{q\in Q}{\min_{p\in T} d(q, p)}$.
Then $d_{\svar{CPD}}(Q_1, T_1)=1$ in Fig.~\ref{fig:distance-example}.
When compared with $d_{\svar{kBCT}}$, $d_{\svar{CPD}}$ is more robust
when erroneous points exist in a trajectory.

\subsection{Trajectory-to-Trajectory}
One weakness in point-to-trajectory measures is that they do not
capture the true ordering of points.
However, many applications of trajectory search expect an ordering
constraint to hold.
That is, the similarity measure should satisfy a 
\textit{local time shifting} \cite{Chen2005a} constraint.

\begin{table}
	\ra{1}
	\centering
	\caption{An overview of existing trajectory-to-trajectory similarity measures.}
	\vspace{-1em}
	\scalebox{0.75}{\begin{tabular}{cccccccc}
		\toprule\addlinespace[0pt]\rowcolor{gray!30}
		 & & & & & \multicolumn{3}{c}{\textbf{Robustness}} \\ \addlinespace[-1pt]\cmidrule{6-8}\addlinespace[-2.5pt]\rowcolor{gray!30}
		\multirow{-2}{*}{\textbf{Type}} & 	\multirow{-2}{*}{\textbf{Measures}} & \multirow{-2}{*}{\textbf{Complexity}} & \multirow{-2}{*}{\textbf{Metric}} & \multirow{-2}{*}{\textbf{\makecell{No Parameter}}} & \textbf{\small GPS Error} & \textbf{\small Sampling Rate} & \textbf{\small Point Shift} \\ \addlinespace[0pt]\midrule
		\multirow{2}{*}{Curve} & Hausdorff \cite{Roh2011} & $\mathcal{O}(n^2)$ & \cmark &\cmark&\xmark&\xmark &\xmark\\ 
		& DFD \cite{Eiter1994} & $\mathcal{O}(n^2)$ & \cmark&\cmark &\xmark&\xmark &\xmark \\\midrule				
		\multirow{2}{*}{Real} & {\dtw} \cite{yi1998efficient} & $\mathcal{O}(n^2)$ & \xmark&\cmark&\xmark&\xmark &\xmark\\ 
		& {\erp} \cite{Chen2004} & $\mathcal{O}(n^2)$ & \cmark&\xmark &\xmark&\xmark &\xmark\\ \midrule
		\multirow{2}{*}{Edit} & {\edr} \cite{Chen2005a} & $\mathcal{O}(n^2)$ & \xmark &\xmark &\cmark&\xmark &\xmark\\ 
		& {\lcss} \cite{vlachos2002discovering} & $\mathcal{O}(n^2)$ & \xmark&\xmark &\cmark&\xmark &\xmark \\ \midrule
		\multirow{2}{*}{Temporal}		& {DISSIM} \cite{Frentzos2007} & $\mathcal{O}(n^2)$ & \xmark&\xmark &\xmark&\cmark &\xmark \\ 
					& {EDwP} \cite{ranu2015indexing} & $\mathcal{O}(n^2)$ & \xmark&\cmark &\xmark&\cmark &\xmark \\ \midrule
			\multirow{2}{*}{Segment} 		& \makecell{{LORS} \cite{Wangsheng2018}, LCRS \cite{Yuan2019}} & $\mathcal{O}(n^2)$ & \xmark&\cmark &\cmark&\cmark &\cmark \\
					& {EBD} \cite{Wangaclustering} & $\mathcal{O}(nlog(n))$ & \cmark&\cmark &\cmark&\cmark &\cmark \\ 					 
		\bottomrule
	\end{tabular}}
	\label{tab:distancemeasures}
		\vspace{-1em}
\end{table}

\subsubsection{Pointwise Measures}
We now provide a taxonomy of similarity measures that can be used to compare and contrast each of them.
As shown in Table~\ref{tab:distancemeasures},  we broadly divide the
most commonly used similarity measures into five categories:
curve-based, real-distance, edit-distance, temporal-aware, and
segment-based.
Each of these will now be described in more detail, and the example
shown in Fig.~\ref{fig:distance-example}(b) will be used to more
concretely illustrate important properties for each measure.

Since a trajectory can also be viewed as a \textit{geometric curve},
Hausdorff distance~\cite{Roh2011} measures the
separation between two subsets of a metric space adversarially, without
imposing an ordering constraint.
Informally, it is the maximum of the distances between each point in a
set $Q$ to the {\em nearest} point in the reference set (trajectory $T$): {\small $d_{\tvar{Hau}}(Q,T)=\max_{q\in Q}{\min_{\ p\in T} {d(p,q)}}$}.\footnote{Note that $d_{\tvar{Hau}}(Q,T)$ is the directed distance from $Q$ to $T$. A more general definition which obeys the symmetry and triangle inequality would be: $\max \{d_{\tvar{Hau}}(Q,T), d_{\tvar{Hau}}(T,Q)\}$, and is metric as shown in Table~\ref{tab:distancemeasures}.}

Discrete Fr\'echet Distance~{\cite{Eiter1994}} (DFD) extends the
Hausdorff distance to account for location and ordering of the points
along the curve, as shown in Equation~\ref{equ:fre}, where $p$ and
$q$ are the head (first) point of $T$ and $Q$, respectively, and
$T_h$ and $Q_h$ are the sub-trajectory excluding the head $p$ and
$q$.
Accordingly, using the example in Fig.~\ref{fig:distance-example}(b),
$d_{\tvar{Hau}}$ and $d_{\tvar{DFD}}$ will return the same value
$3.1$, since the matching shown in the figure not only finds the nearest
neighbor for each point but also obeys local time shifting (the order
constraint discussed above).
If any two points in $T_5$ swap locations, $d_{\tvar{Hau}}$ does not
change while $d_{\tvar{DFD}}$ could.

\vspace{-1em}
\begin{equation}\small
\label{equ:fre}
d_{\tvar{DFD}}(Q,T)=\begin{cases}d(p, q),\ \text{if}\ Q=\{q\}\ and\ T=\{p\}\\
\infty,\ \text{if}\ Q=\emptyset\ or\ T=\emptyset\\
\max\{d(q, p), \min\{d_{\tvar{DFD}}(Q_h,T_h),\ d_{\tvar{DFD}}(Q, T_h), d_{\tvar{DFD}}(Q_h, T)\}\},\ \text{otherwise}
\end{cases}
\end{equation}

\medskip

Since a trajectory can be also viewed as a time series, many similarity
measures originally designed for time series search can be leveraged
in trajectory related problems~{\cite{Zheng2015b}}.
Time-series measures which have been applied to trajectory data
include Dynamic Time Warping (\dtw) \cite{yi1998efficient}, Longest
Common Subsequence (\lcss)
\cite{vlachos2002discovering}, and Edit Distance for Real
sequences (\edr) \cite{Chen2005a} and Edit distance with Real Penalty
{(\erp)}~\cite{Chen2004}.
{\dtw} computes the distance based on the sum of minimum distances,
instead of choosing the maximum as in Discrete Fr\'echet Distance, as shown in
Equation~\ref{equ:dtw}.
To compute {\dtw} for Fig.~\ref{fig:distance-example}(b), every
possible pairing of points in $T_5$ and $Q_3$ are compared in order,
and the sum of the minimum of each pairing
{\small $d_{\tvar{DTW}}(Q_3,T_5)=1.2+3+1+2.2+3.1=10.5$}.

\vspace{-0.5em}
\begin{equation}\small
\label{equ:dtw}
d_{\tvar{DTW}}(Q,T)=\begin{cases}d(p, q),\ \text{if}\ Q=\{q\}\ and\ T=\{p\}\\
\infty,\ \text{if}\ Q=\emptyset\ or\ T=\emptyset\\
d(q, p)+\min\{d_{\tvar{DTW}}(Q_h,T_h),\ d_{ \tvar{DTW}}(Q, T_h),\ d_{\tvar{DTW}}(Q_h, T)\},\ \text{otherwise}
\end{cases}
\end{equation}

\medskip

Instead of computing the real distance, computing an edit distance (0
or 1) can be more robust for noisy data as outliers can heavily
impact {\dtw}-based comparisons.
To judge whether two points are matched, edit-distance based measures
set a range threshold $\tau$, i.e., $\var{match}(p, q)=\var{true},\ \text{if}\ |p_x- q_x|\le\tau\ \text{and}\ |p_y-q_y|\le\tau$, where $| |$ denotes the absolute value, and $p_x$
and $p_y$ denote the latitude and longitude of point $p$,
respectively. 
Now \lcss and \edr can be defined as shown in Equation~\ref{equ:lcss}
and Fig.~\ref{equ:edr}.
The main difference between \lcss and \edr is that \lcss 
measures the similarity between two trajectories, while \edr 
measures the {\em dissimilarity}.
For example, consider $\tau=0.5$ for each of these in
Fig.~\ref{fig:distance-example}(b).
The result is no matching point for $d_{\scriptscriptstyle
\tvar{LCSS}}(Q_3,T_5)=0$ and $d_{\scriptscriptstyle
\tvar{EDR}}(Q_3,T_5)=4$.
If $\tau=5$, then every point can be matched, and
$d_{\scriptscriptstyle
\tvar{LCSS}}(Q_3,T_5)=4$ and $d_{\scriptscriptstyle
\tvar{EDR}}(Q_3,T_5)=0$.
Hence, both measures are sensitive to the hyperparameter
$\tau$.

\vspace{-0.5em}
\begin{equation}\small
\label{equ:lcss}
d_{\scriptscriptstyle \tvar{LCSS}}(Q,T)=\begin{cases}0,\ \text{if}\ Q=\emptyset\ \text{or}\ T=\emptyset\\
1+d_{\scriptscriptstyle \tvar{LCSS}}(Q_h, T_h),\ \text{if}\ \var{match}(p,q)=\var{true}\\
\max\{d_{\scriptscriptstyle \tvar{LCSS}}(Q, T_h),\ d_{\scriptscriptstyle \tvar{LCSS}}(Q_h, T)\},\ \text{otherwise}
\end{cases}
\end{equation}


\vspace{-0.5em}
\begin{flalign}\small
\label{equ:edr}
d_{\scriptscriptstyle \tvar{EDR}}(Q,T)=\begin{cases}
|Q|\ or\ |T|,\ \text{if}\ T=\emptyset\ or\ Q=\emptyset\\
\min\{d_{\scriptscriptstyle \tvar{EDR}}(Q_h,T_h), d_{\scriptscriptstyle \tvar{EDR}}(Q, T_h)+1,\ d_{\scriptscriptstyle \tvar{EDR}}(Q_h, T)+1\},\ \text{if}\ \var{match}(p,q)=\var{false} \\
\min\{1+d_{\scriptscriptstyle \tvar{EDR}}(Q_h,T_h), d_{\scriptscriptstyle \tvar{EDR}}(Q, T_h)+1,\ d_{\scriptscriptstyle \tvar{EDR}}(Q_h, T)+1\},\ \text{otherwise}
\end{cases}
\end{flalign}

\medskip

Since these measures (\dtw, \lcss, and \edr) do not obey the triangle inequality, which is
an essential requirement for \textit{metric} space pruning, all of the
time-series similarity measures are \textit{non-metric} except for
{\erp} \cite{Chen2004}.
In \erp, a point $g$ can be any fixed point in a metric space, and is used as
the reference origin point $g=\{0,0\}$.
Changing $g$ can also change
distance scores similar to \lcss, which can result in 
{\knn}-based algorithms being non-deterministic. 

\vspace{-1em}
\begin{flalign}\small
\label{equ:erp}
d_{\scriptscriptstyle \tvar{ERP}}(Q,T)=\begin{cases}\sum_{p\in T}d(g, p),\ \text{if}\ Q=\emptyset\\
\sum_{q\in Q}d(q, g),\ \text{if}\ T=\emptyset\\
\min\{d_{\scriptscriptstyle \tvar{ERP}}(Q_h,T_h)+d(q, p),d_{\scriptscriptstyle \tvar{ERP}}(Q, T_h)+d(g, p),d_{\scriptscriptstyle \tvar{ERP}}(Q_h, T)+d(q, g)\},\ \text{otherwise}
\end{cases}
\end{flalign}

\subsubsection{Temporal-aware Pointwise Similarity}
The aforementioned six measures are the representative of commonly
deployed similarity measures for spatial-only trajectory applications.
In addition to spatial information, temporal information is also an
important factor in accurate sample rate calibration.\footnote{{Actually all aforementioned spatial-only measures can be easily extended to handle the spatio-temporal case, by performing two separate distance calculations on the spatial and temporal features, and then combining the score by using a balancing factor $\alpha$.}}
\citet{Frentzos2007} proposed a measure called {\footnotesize \textsf{DISSIM}} to compute the
dissimilarity: $d_{\svar{DISSIM}} = \int_{t_1}^{t_n}d(Q_t, T_t)dt$. Here, $Q_t$ and $T_t$ denote the points of $Q$ and $T$ at timestamp
$t$ in the range $[t_1, t_n]$.
The core idea is to integrate time w.r.t.\ Euclidean distance for
trajectories occurring in the same period.
{\footnotesize \textsf{DISSIM}} can resolve various sampling rate problems commonly
encountered through integration, but can also be computationally
expensive.
To alleviate this issue, {\footnotesize \textsf{EDwP}}
\cite{ranu2015indexing} calibrates trajectory manipulations by adding
or removing points to align the sampling rate between any two
trajectories, and computes similarity as described for {\edr}.

\subsubsection{Segment-based Similarity}
\label{sec:segmentsim}
Temporal-aware pointwise methods can compute similarity
in much richer trajectory data, but results often have precision
issues unless sample-rate calibration is applied.
Another alternative is to convert trajectories to segments.
This approach has been shown to reduce sample mismatch effects
as well as reducing the complexity of the similarity computations
applied.
\citet{Tiakas2009} was among the first to propose this solution for
road networks.
The approach uses the sum of the distances between nodes in the road
network that contribute to the final distance, but the trajectory
pairs must have the same length in order to guarantee point-to-point
matching.
To circumvent the length constraint, \citet{Mao2017} proposed a
related solution which used \dtw after converting pointwise
trajectories to paths on a road network.
In both approaches, the similarity is still computed based on
the end-points of road segments, so map matching is only applied once to
clean trajectory data and to align sampling rates, but it still needs
to compute the Euclidean distance between nodes which can be computationally
expensive even is small data sets.

To reduce the costs in when computing the distances,
\citet{Wangsheng2018} proposed the use of
\textit{longest overlapped road segments} (\lors). 
This measure was inspired by {\lcss} and adapted to leverage the
properties inherent in map-matched data.
{\lors} does not compute the Euclidean distance between points to
determine if they adhere to a threshold distance constraint.
Instead, overlapping segments between trajectories are identified
first, and then reused to compute the similarities.

\vspace{-0.5em}
\begin{equation} \small
\label{equ:lors}
d_{\svar{LORS}}(Q,T)=
\begin{cases}
0,\ \ \ \ \text{if}\ Q\ \text{or}\ T\ \text{is empty} \\
|e_{1m}|+d_{\svar{LORS}}(Q_h,T_h), 
\ \ \ \ \text{if}\ e_{1m}=e_{2x}\ \\
\max(d_{\svar{LORS}}(Q_h,T),d_{\svar{LORS}}(Q,T_h)),\text{otherwise}
\end{cases}
\end{equation}

\smallskip

Here, the inputs are $Q=(e_{11},e_{12},...,e_{1m})$ and
$T=(e_{21},e_{22},...,e_{2x})$, where $e_{1m}$ are the end edge of
trajectory $Q$, and $|e_{1m}|$ is the travel length of a graph edge
$e_{1m}$.
\citet{Yuan2019} further extended and normalized {\lors} to account for 
trajectory length effects in $|Q|$ and $|T|$.
The resulting measure LCRS can be defined as: {\small
$d_{\svar{LCRS}}(Q,T) =
\frac{d_{\svar{LORS}}(Q,T)}{|Q|+|T|-d_{\svar{LORS}}(Q,T)}$}.
By modeling network-constrained trajectories as strings in a similar
manner, {\citet{Koide2020}} proposed a generalized metric called
\textit{weighted edit distance} ({\small WED}), which supports
user-defined cost functions that can be used with several important
similarity functions such as {\erp} and {\edr}.

Both {\lors}, {\small LCRS}, and {\small WED} satisfy the local time
shift constraint, and have a quadratic complexity using dynamic
programming.
\citet{Wangaclustering} later simplified the computational 
costs of \lors further.
The key insight was to exploit ordered integer intersection (through
finger search), where ordered integers represented unique IDs
assigned to road segments during map-matching.
The resulting method called {\em edge-based distance}-EBD: {\small
$d_{\svar{EBD}}(Q, T) = \max(|Q|,|T|)-|Q\cap T|$}, was shown to have
comparable precision to {\lors} and was also highly scalable in
practice.

\subsection{Complexity and Robustness}

Table~\ref{tab:distancemeasures} compares characteristics of
similarity measures based on four different properties: complexity,
metricity, parameter independence, and robustness.
Since parameters such as $\tau$ in \lcss and \edr can be easily
observed from their definitions, we will elaborate the other three
characteristics next.

\myparagraph{Complexity} Many trajectory-based similarity measures
were designed to allow search to be more resilient to variance
produced by local time shifting.
However, computing optimal distances with ordering constraints
generally requires dynamic programming solutions with quadratic
complexity ($\mathcal{O}(n^2)$) as well as the associated space
overheads.

\begin{figure}
	\begin{minipage}{0.67\textwidth}
		\centering
		\includegraphics[width=1\linewidth]{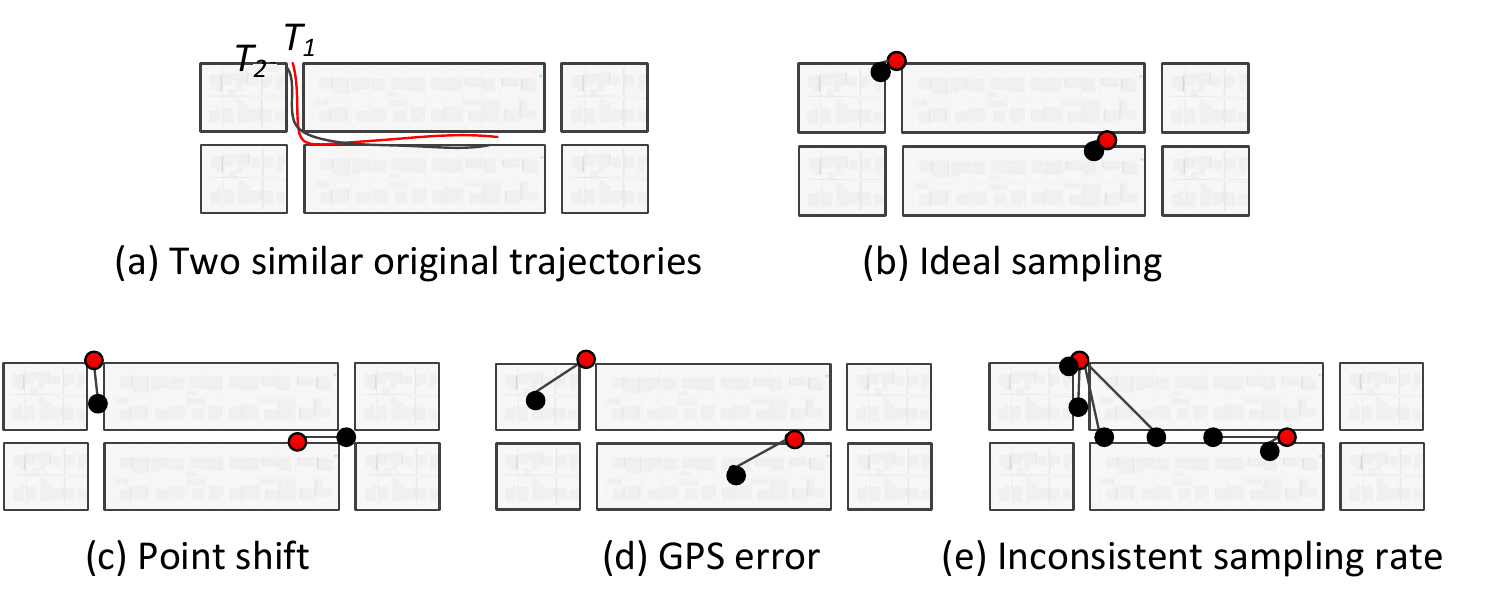}
		\vspace{-1.5em}
		\caption{A similarity measure that is sensitive to point shift, error, and sampling rate.
			}\vspace{-1.2em}
 \label{fig:robustness}
	\end{minipage}\hfill
	\begin{minipage}{0.3\textwidth}
		\centering
		\includegraphics[width=1\linewidth]{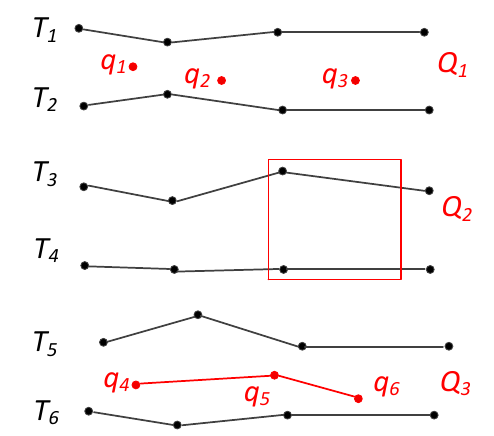}
		\vspace{-1.5em}
		\caption{Three kinds of spatial-only trajectory queries.}
		\vspace{-1.2em}
 \label{fig:queries}
	\end{minipage}
\end{figure}

\myparagraph{Metricity} Fully metric compliance is often crucial for
a similarity measure, as obeying the triangle inequality provides
both theoretical and practical advantages.
Here, we use an example to illustrate the concept.
Consider the computation of the well-known metric similarity measure,
Euclidean distance.
Given a query point $q$ with $p_1$ and $p_2$ as the two next
candidates for a \knn query, $d(q,p_1)$ is computed.
Before distance is computed for $p_2$, the lower and upper bound of
$d(q,p_2)$ can be estimated based on $d(q,p_1)$ and $d(p_1, p_2)$ as:
{\small $ |d(q, p_1)-d(p_1, p_2)|<d(q,p_2)<d(q,p_1)+d(p_1, p_2) $},
where $d(p_1, p_2)$ can be precomputed off-line for any query.
The lower bound can be compared with $d(q, p^{'})$ where $p^{'}$ is
currently the best nearest neighbor candidate.
If the bound is greater than the current best result, we can discard
$p_2$ directly as it can never replace $p^{'}$.

\myparagraph{Robustness}
In addition to complexity and metricity, many similarity measures are
sensitive to noise, sampling rate, and point shifting (points are
shifted on the path where the trajectory lies).
The existence of noise due to GPS errors or other common sensor data 
collection approaches result in more
(or fewer) sampled points, point shifting, and other quality issue.
\citet{Wang2013a} systematically compared the robustness of 
six common similarity measures in a real-world trajectory
dataset, and a complete version \cite{Su2019b} was published later
with additional point-based measures.
However, the most recent segment-based measures
\cite{Wangsheng2018,Yuan2019} were not covered.

Fig.~\ref{fig:robustness} illustrates several common robustness
problems encountered in trajectory data.
$T_1$ and $T_2$ are two vehicle trajectories in a road network that
overlap.
After sampling, a set of points represents each one, with perfect
data resulting in (b).
In reality, sampling rates can vary, resulting in a point shift
(c).
GPS errors (d) can also occur in sampled points that are not on the
same road, and inconsistent sampling rates (e) lead to a different
number of points being produced for the two trajectories.
Each of the above issues can lead to distance scores greater than
those found in error free data, and can even result in incorrect
solutions for certain query types.
Robust similarity measures are therefore highly desirable.

\section{Trajectory Search and Join}
\label{sec:spatialonly}

\subsection{Spatial-only Trajectory Search}
Fig.~\ref{fig:queries} shows three exemplar queries $Q_1$, $Q_2$,
$Q_3$ for the spatial-only trajectories $T_1$ to $T_6$.
Given three points ($q_1$, $q_2$, $q_3$), $Q_1$ finds the best
connected trajectory.
Among all six trajectories, $T_1$ and $T_2$ are both possible 
solutions.
For a pointwise similarity search as shown in $Q_1$,
the similarity measure used should ideally be able to reliably
distinguish between the two candidates.
Given a target range (red box), $Q_2$ finds all trajectories covered
(partially or fully).
Valid results are $T_3$ and $T_4$, and can not be computed using any
of the similarity measures shown.
In comparison, if $Q_3$ is the query for the top-$k$ search, a
similarity measure must to be defined that can capture all of the
properties of $Q_3$, including points and ordering.

\begin{table*}
	\centering
	\ra{0.8}
\caption{{An overview of common operations applied to trajectory and
point data.}}
\vspace{-1em}
	\label{tab:methods}
	\scalebox{0.76}{\begin{tabular}{ccccccc}
			\toprule\addlinespace[0pt]\rowcolor{gray!30}
			\textbf{Data}&\textbf{Type}&\textbf{Query} & \textbf{Input} & \textbf{Output} & \textbf{Measure}& \textbf{Index} \\\addlinespace[-2pt]
			\midrule
			\multirow{9}{*}{\makecell{Spatial\\-only}}	&\multirow{2}{*}{Basic}& Range \cite{SanduPopa2011,Krogh2016} & range & trajectories& N/A & R-tree \\
			&& Path \cite{Krogh2016,Koide2015} & path & trajectories& N/A & R-tree \\
			\cmidrule{2-7}
			&\multirow{4}{*}{$k$NN}&\knn~\cite{Guttman1984,beckmann1990r} & point & points & Euclidean & R-tree \\
			&&$k$NNT \cite{vlachos2002discovering,Chen2005a,Shang2018}& trajectory & trajectories& LCSS, DTW & R-tree \\
			&&\kbct \cite{Chen2010a,Tang2011}& points & trajectories& Aggregate & R-tree \\
			\cmidrule{2-7}
			& \multirow{2}{*}{\makecell{Reverse\\ $k$NN}} & \rknn \cite{Tao2004}& \makecell{point} & points& Euclidean & R-tree \\
			& & \rknnt~\cite{Wang2018c}& {trajectory} & trajectories & CPD & N/A \\
			\cmidrule{2-7}
			& \multirow{2}{*}{Clustering} & Density \cite{Lee2007}& \makecell{thresholds} & paths& N/A & N/A \\
			& & Partition \cite{Wangaclustering}& \makecell{$k$} & trajectory& EBD & Inverted index \\ 	\cmidrule{2-7}
			 & \multirow{2}{*}{Join} & \makecell{similarity join} \cite{Ta2017,Yuan2019}& threshold & pairs& Normalized CPD & signature \\
			 & & \knn join \cite{Fang2016}& $k$ & pairs& CPD & Grid-index \\
			\midrule
			\multirow{3}{*}{\makecell{Spatial\\-textual}}	& &\textbf{KS} \cite{Broder2003,Ding2011a} & keywords & points & TF-IDF & Inverted list \\
		& \multirow{2}{*}{Top-$k$} &\tksk \cite{zhang2014processing,Li2011} & \begin{tabular}{c}point, keywords\end{tabular} & points & Aggregate & IR-tree \\
			& &\tkstt \cite{Shang2012a,Zheng2013c}& \begin{tabular}{c}points, keywords\end{tabular} & trajectories & Aggregate & \begin{tabular}{c}Grid-index, \\Inverted list\end{tabular} \\
			\bottomrule
	\end{tabular}}
	\vspace{-1em}
\end{table*}

Table~\ref{tab:methods} compares and contrasts several common
trajectory queries by input, output, similarity measure, and most
appropriate index.
The ``Index'' column shows the preferred indexing approach.
Many other alternatives exist.
A more comprehensive analysis and discussion of indexing solutions is
in Section~{\ref{sec:spatialindexing}}.
Query types range from basic trajectory search, top-$k$ trajectory
similarity search (spatial-only and spatial-textual), to more complex
operations such as reverse $k$ nearest neighbors search, and
trajectory clustering.
We now discuss each of these query operations in detail along with
several indexing techniques which have been proposed to accelerate
performance.

\subsubsection{Basic Trajectory Search}
Basic trajectory search includes three classic query
formulations.
The most basic is a {\em Range Query}
({RQ})~{\cite{SanduPopa2011,Krogh2016,Song2014,Roh2011}} which finds
all (sub-)trajectories located in a spatial or temporal region.
Range queries has many applications in traffic monitoring such as
returning all vehicles at a road intersection.
The second is a {\em Path Query} ({PQ}) which retrieves the
trajectories that contain any edge of the given path query.
The third is a {\em Strict Path Query}
({SPQ})~{\cite{Krogh2016,Koide2015,SanduPopa2011}} which
finds all trajectories that traverse an entire path from
beginning to end.
Interestingly, path queries and strict path queries share many common
properties with disjunctive (OR) and conjunctive (AND) Boolean
queries~{\cite{Christopher2009}}.

\begin{definition}\textbf{(Range Query)}
Given a trajectory database $D = \{T_1,
\ldots, T_{|D|}\}$ and query rectangular region $Q_r$, a range query
	retrieves the trajectories:
$\avar{RQ}(Q_r)=\{T\in D|\exists p_i \in T(p_i\in Q_r)\}$.
\end{definition}

\begin{definition}\textbf{(Path Query)}
Given a $Q_p$ which is a path in $G$, a path query retrieves the
trajectories {$T$ that pass through at least one edge $e_j$ of $Q_p$}: $
\avar{PQ}(Q_p)=\{T \in D|\exists e_i \in {T}, e_j \in Q_p (e_i = e_j)\}
$.
\end{definition}

\begin{definition}\textbf{(Strict Path Query)} 
Given a $Q_p$, a strict path query retrieves the trajectories whose edges can all be found in $Q_p$: $
\avar{SPQ}(Q_p)=\{T\in D|\exists i,j(T_{ij}=Q_p)\}
$, where $T_{ij}=\{e_i,e_{i+1},\cdots,e_j\}$ is the
sub-trajectory of $T$.
\end{definition} 

\subsubsection{$k$ Nearest Neighbors Query}

To find a relevant subset from a large set of objects for a given query object, $k$ nearest neighbor queries have been widely applied in spatial databases. For trajectories, the query can be either a trajectory or a set of points.

\begin{definition}\textbf{(k Nearest Neighbors Query over Trajectories)}
Given a trajectory database $D = \{T_1, \ldots, T_{|D|}\}$ and query
$Q=\{q_1, q_2, \cdots, q_{|Q|}\}$, a $k$ Nearest Neighbors Query
($\knn(Q)$) retrieves a set $D_s \subseteq D$ with $k$ trajectories
such that: $\forall T \in D_s, \forall T^{'} \in D - D_s, d(Q,T) <
d(Q,T^{'})$.
\end{definition}

\myparagraph{Search by Trajectory}
Given a query trajectory $Q$, a $k$ Nearest Neighbor Trajectories
Query aims to find the $k$ most similar/nearest trajectories to
$Q$, based on a given trajectory similarity measure \cite{Xie2017a,Wangsheng2018}.
Such a query can be used to find the most similar trips in traffic
flow analysis.
For the special case when the trajectory is a single point, which is
also known as \knn search (we use \textit{kNNT} to denote trajectory
search), with the default similarity being Euclidean distance.

\myparagraph{Search by Points}
There has also been previous work targeting search on spatial-only
trajectory data where the input query is a set of
points~\cite{Chen2010a,Tang2011,Qi2015}.
\citet{Chen2010a} initially proposed the problem of querying over a
set of points with spatial-only trajectories.
They proposed incremental expansion using an R-tree
to prune candidate points from consideration.
They referred to the approach {\em Incremental K Nearest Neighbors
(IKNN)}.
To optimize the IKNN algorithm,
\citet{Tang2011} devised a qualifier expectation measure that ranks
partially matched candidate trajectories.
The approach accelerates query processing significantly when
non-uniform trajectory distributions and/or outlier query locations
exist in the solution space.

\subsubsection{Reverse $k$ Nearest Neighbor Query}
\label{sec:rnns}
Instead of searching the most relevant objects to a query object, {\em Reverse $k$ Nearest Neighbors} ({{\rknn}}) queries attempt to locate objects which will take the query as one of their $k$ nearest neighbors. Formally, \rknn is defined as below:

\begin{definition}\textbf{(Reverse k Nearest Neighbor)}
Given a set of points (or trajectories) $D$ and a query point (or trajectory) $Q$, a $\rknn(Q)$
retrieves all objects $\tran \in D$ {that take $Q$ as $k$NN}, i.e., $\forall\tran$,
$Q\in\textbf{\knn}(\tran)$.
\end{definition}

\rknn queries for spatial
point data have attracted considerable attention in the research
community~{\cite{Tao2004}} as it can be used to solve a
wide variety of industry-relevant problems.
An {\rknn} query aims to identify all (spatial) objects that have a
query location as a $k$ nearest neighbor.
Important applications of the {\rknn} query include resource allocation \cite{Wang2018c} and profile-based marketing \cite{Zhang2018c}.
For example, {\rknn} queries can be used to estimate the number of
customers a new restaurant would attract based on location, and was
initially referred to as a \textit{bichromatic} \rknn based on the
dual-indexing solution proposed to solve the problem~\cite{Tao2004}.

Instead of querying a set of static points,
\citet{Cheema2012} proposed the \textit{continuous reverse nearest
	neighbors} query to monitor a moving object.
The goal is to find all static points that contain the moving object
as a $k$ nearest neighbor.
This approach targets a single point rather than a commuter
trajectory of multiple-points, which was later solved in \rknnt~\cite{Wang2018c}.

\myparagraph{Reverse Trajectory Search}
The ``Reverse $k$ Nearest Neighbor Query over Trajectories''
(\rknnt~\cite{Wang2018c}), and can be defined as follows.
Given a trajectory dataset $\mathcal{D}_T$ and a set of routes
$\mathcal{D}_R$ (also defined as trajectory) and a candidate point
set $Q=\left( o_1,o_2,\ldots,o_m \right)$ as a query, \rknnt~returns all
the trajectories in $\mathcal{D}_T$ that will take $Q$ as $k$ nearest
routes using a point-to-trajectory similarity measure.
The main application of \rknnt~is to estimate the capacity of a new
bus route and can be further used to plan a route with a maximum
capacity between a source and destination, which was defined as
a \maxrknnt~in \cite{Wang2018c}.
The main challenge in solving \rknnt~is how to prune the
trajectories which cannot be in the results without explicitly
accessing the whole dataset $\mathcal{D}_T$.
All of the pruning methods commonly used for {\rknn} may work for
trajectories.
For example \textit{half-space} pruning~\cite{Tao2004}, which can prune
an area by drawing a perpendicular bisector between a query point and
a data point will not work.
\citet{Wang2018c} proposed the use of an R-tree of the trajectory
routes (an example can be seen in Fig.~\ref{fig:R-tree-indexing}).
By drawing bisectors between the route points and the query, an area
can be found where all trajectory points contained in a region can
not have the nearest route for the query.


\subsection{Spatial Indexing Algorithms}
\label{sec:spatialindexing}
Trajectory indexing is commonly applied to improve efficiency in 
trajectory related search problems.
Trajectory data is processed off-line to improve scalability and
prune the search space.
Existing approaches to trajectory indexing have two components:
indexable point data and mapping tables.

\begin{figure}
	\centering
	\includegraphics[height=4cm]{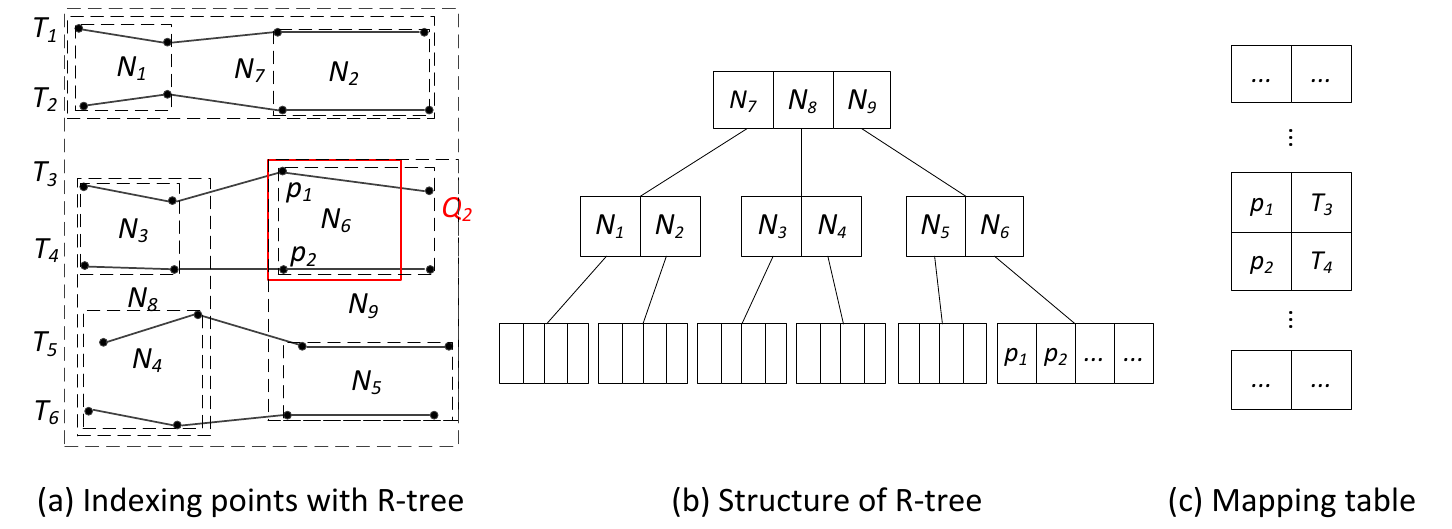}
	\vspace{-0.5em}
	\caption{Indexing spatial-only trajectories using an R-tree and a mapping table.}
	\vspace{-1.5em}
	\label{fig:R-tree-indexing}
\end{figure}

\myparagraph{Point Indexing}
Space-efficient index representations and processing frameworks are
crucial for trajectory data, and the majority of trajectory
search solutions~\cite{Chen2010a} rely on an
R-tree~{\cite{Guttman1984}}, which store all points from the raw
trajectories, and are historically the dominant approach deployed for
spatial computing applications.
As shown in Fig.~\ref{fig:R-tree-indexing}, trajectories are
decomposed to points first, then an R-tree is used to index each 
point.
A mapping table identifies the trajectory that contains each point.
Since trajectory datasets such as 
T-drive~{\cite{Yuan2010}} can easily contain millions of
points, the R-tree must manage an enormous number of maximum bounding
rectangles (MBR), which have prohibitive memory costs in practice.
So, traditional methods employed for spatial pruning can be
infeasible on raw trajectory data.
Simpler Grid-index solutions can sometimes be more appropriate in
such scenarios~{\cite{Zheng2013c,Wang2021}}.
Nevertheless, the core problem is compounded by the fact that many of
the similarity measures proposed for trajectory search are
non-metric.

\myparagraph{Mapping Table}
A mapping table is used to map points to trajectories
\cite{Chen2010a,Zheng2013c}.
After searching for a point, the mapping table identifies the
trajectory containing the point.
Ensure that the mapping is unambiguous, every point has a unique
identifier ranging from $[1, |\mathcal{D}.P|]$, each trajectory
has a unique identifier ranging from $[1, |\mathcal{D}|]$,
$|\mathcal{D}.P|$ and $|\mathcal{D}|$ are the number of points and
trajectories.
With the mapping table in Fig.~\ref{fig:R-tree-indexing}(c), we can know that $p_1$ is a point of $T_3$ and $p_2$ belongs to $T_4$.

\myparagraph{Pruning Mechanism}
For range queries, most existing approaches employ MBR-based pruning,
as MBR intersection with a query trajectory can be used to
effectively and efficiently prune the search space.
If a node intersects with a query range, the object covered
can be removed from consideration.
For example, the nodes $N_7$ and $N_8$ in
Fig.~\ref{fig:R-tree-indexing}(a) do not intersect with $N_2$, so
they can be eliminated from consideration, leaving only $N_9$.
For \knn trajectory search, there are two common pruning techniques:
\textbf{early termination} and \textbf{early abandoning}.
Early termination is initially proposed in the threshold algorithm
\cite{Fagin2003}.
Given an index, the algorithm scans a sorted list, and early
terminates when all remaining items exceed some precomputed upper
bound.
It is similar to pruning in {\knn} search with an R-tree where
the minimum distance between the query and a node in
the index can be used to determine if the distance for the
items contained must be computed.
In contrast to early termination which can prune many items
in a single operation, early abandoning must consider each
item but exploits a bound to minimize the number of expensive
computations applied during processing.
This technique is commonly used to accelerate time series similarity
search and {\knn} search~\cite{vlachos2002discovering}.
By estimating the upper bound of the similarity score and comparing
it to the $k$-th item in a max-heap (for top-$k$ search for example),
an algorithm can determine if a full similarity computation must be
made for the item under consideration.

\subsection{Trajectory Join}
\begin{definition}\textbf{(Trajectory Join)}
	\label{def:join}
	Given two sets of trajectories $S_1$ and $S_2$, and a similarity
	threshold $\epsilon$ (or $k$ in \knn search), a trajectory join
	operation will return all pairs of trajectories $T_i$ and $T_j$ from
	$S_1$ and $S_2$ with a similarity that exceeds $\epsilon$ (or $T_i\in
	\knn(T_j)$).
\end{definition}

The main applications of trajectory joins are in trajectory data
cleaning, near-duplicate detection, and carpooling.
For example, given a driver trajectory database and a rider
trajectory database, a trajectory join will match riders with similar
trajectories to drivers.
Trajectory join operations can be divided into two categories: trajectory similarity joins~\cite{Ta2017} and
trajectory $k$ Nearest Neighbors joins~\cite{Fang2016}.
A simple baseline is to conduct a similarity search or {\knn} search
for each trajectory.
Given the quadratic complexity of this solution, scaling can be
problematic.

To reduce the complexity, many
different indexing techniques can be used.
\citet{Bakalov2008} considered the moving object trajectory similarity
join, which is a pointwise join in a specific timestamp, and
similarity is computed between points and not trajectories.
\citet{Ta2017} proposed a signature-based solution to filter
trajectory pairs before distance computation, 
where the signature is built for each trajectory to help prune, 
based on its crossed and neighbor grids.
\citet{Shang2017} explored the spatial-temporal trajectory similarity
join problem for road networks.
A two-step algorithm is applied to each trajectory: expansion and
verification, similar to trajectory similarity search in
Section~\ref{sec:spatialonly}.

\myparagraph{Distributed Trajectory Similarity Join} Distributed computing is often applied to solve the
trajectory join problem due to the high computational costs of the
distance computation.
\citet{Fang2016} proposed a distributed trajectory \knn join method using MapReduce.
Their solution used Hadoop, and the algorithm can also run on a
single machine.
\citet{Shang2018} computed distributed trajectory similarity joins
using Spark, where a global index is deployed over multiple computer
nodes and used to prune the search space.
\citet{Yuan2019} also explored the distributed trajectory similarity
join problem using similar techniques to \citet{Shang2018}, but 
applied the similarity measure {LCRS}~described in Section~\ref{sec:segmentsim} to compute.

\subsection{Spatial-Textual Trajectory Search}
\label{sec:spatialtextual}
Spatial-textual trajectory search incorporates text and keywords
to the data, and relevant trajectories have more than one
dimension of similarity that must be considered.
For example the keywords ``\textit{seafood}'', ``\textit{coffee}''
and ``\textit{swimming}'' might be contained in a query and a
trajectory.

\subsubsection{Top-$k$ Spatial-Textual Trajectory Search}
\label{sec:literature-etq}

\myparagraph{Similarity Measures}
In contrast to spatial-only similarity measures, spatial-textual
measures for an activity trajectory depend on both spatial and
textual components.
Linear combinations of two distinct similarity measures is a common
solution for spatial-textual data~\cite{Cong2009a,zhang2014processing}.
The spatial distance is combined with TF$\cdot$IDF similarity for the
text, with a user-defined weight between the two to balance
importance based on the target application.
Such a measure was also extended by \citet{Shang2012a}, where the
spatial-distance is computed in a manner similar to the \textit{$k$
Best Connected Trajectories query} (\kbct).
The textual similarity was then computed using simple keyword
intersection.
Since the text relevance is
computed as a part in the final score, it is more flexible in real
applications.
Conversely, a conjunctive approach \cite{Zheng2013c,Cong2012}
tightens the constraint, requiring that the result trajectory should
contain all of the query keywords, and results are ordered by spatial
distance.

\myparagraph{Pointwise Search with Keywords} Several recent papers
have explored the problem of spatial-textual trajectory search in a
range of different scenarios.
\citet{Cong2012} proposed a sub-trajectory based solution to find the
minimum travel distance for a single query point using a Boolean
keyword match.
\citet{Shang2012a} presented a disjunctive solution for 
multiple points, where the distance is measured between
points, but keywords are assigned to an entire trajectory instead of
each point.

\citet{Zheng2013c} attached the keywords to a specific point in the
trajectory to allow finer-grained matching of textual information.
However, their work only supported conjunctive text matching, so
results must contain all query keywords, simplifying the more
general keyword search problem.
To handle the case when users do not have a preferred locations
or exact text matching should not be guaranteed,
\citet{Zheng2015} proposed an approximate query solution over
trajectories that use a set of keywords, and the similarity is
measured as the travel distance, just as
\citet{Cong2012} did.
A special case of this problem is when the query contains only
keywords, and is known as a keyword-only search (\textbf{KS})
\cite{Broder2003}.
Another special case is when the trajectory data and query are a
single point, and is known as Top-$k$ Spatial Keyword Search (\tksk)
\cite{zhang2014processing}.

\subsubsection{Boolean Range and Top-$k$ Query}\citet{Hariharan2007} utilized the Boolean range queries to find all objects
containing query keywords, all of which must be located within a bounded
range.
Boolean Top-$k$ queries \cite{DeFelipe2008a} find the $k$ nearest
objects using conjunctive Boolean constraints on the text, i.e., it
conducts a {\knn} search over all the points that contain all the
keywords from query.
\citet{HanWZZL15} investigated Boolean range queries on trajectories,
and considered spatial, temporal, and textual information in the
solution, and an index based on an octree and inverted index is proposed to
answer their proposed query. 


\subsubsection{Pruning with an Inverted Index}
Inverted indexes~\cite{Zobel2006} are widely used to efficiently
manage text information for spatial-textual problems.
Related algorithms and data structures for inverted indexes have
benefited from many years of practical improvements to support web
search and related applications, making them an obvious candidate
to manage textual components in heterogeneous data collections.
A unique object identifier is mapped to every keyword contained,
and an inverted list of all objects containing each keyword
can only represent existence, or contain auxiliary data such as
the frequency of the keyword in the object text.
The auxiliary data can also be a precomputed similarity score or
even offsets for phrase reconstruction.
Most prior work~\cite{zhang2014processing} attaches
inverted lists to nodes in an existing tree-based spatial index
(a Grid-index for example) in order to support spatial-textual
similarity computations during tree traversal.

\section{Trajectory Clustering and Classification}
\label{sec:clustering}
\subsection{Clustering Large-Scale Trajectories}
Clustering similar trajectories to produce representative exemplars can be a powerful visualization tool to track the mobility
of vehicles and humans.
It has been investigated in many different applications, such as
spatial databases
\cite{Hung2015, Agarwal2018, Lee2007}, data mining
\cite{Pelekis2009}, transportation \cite{Wu2015}, and visualization
\cite{Ferreira2013}.
Clustering is most often applied to spatial-only
trajectories, with prior work on spatial-textual trajectory
clustering being relatively rare.
Trajectory clustering can be broadly divided into two categories:
Partition-based:
\cite{Chan2018,Ferreira2013,Yao2017,Pelekis2009,Gudmundsson2015,Hung2015} and
Density-based:
\cite{Lee2007,Han2015a,Li2010b,Agarwal2018,Andrienko2009}.
Both the partition and the density-based trajectory clustering require
extensive similarity computations, with the only distinction being if
it is computed for whole trajectories or using only sub-trajectories.

{
\subsubsection{Partition-based Clustering}
Given a set of trajectories, partition-based clustering divides
trajectories into a limited number ($k$) of groups (clusters).
The similarity measure (Section~\ref{sec:distance}) and parameter $k$
are selected a priori depending on the use case.

\begin{definition}\textbf{(Partition-based Trajectory Clustering)}
Given a set of trajectories $\{{T}_1,{T}_2,\cdots,{T}_n\}$,
partition-based clustering aims to partition $n$ trajectories
into $k$ ($k\le n$) clusters $S=\{S_1, S_2, \cdots,S_k\}$ by minimizing
the objective function: $O=\argmin_{S}{\sum_{j=1}^{k}{\sum_{T_i\in
S_j}{d(T_i,\mu_j)}}}$.
\end{definition}

Here, each cluster $S_j$ has a centroid trajectory or path $\mu_j$,
and $d(T_i,\mu_j)$ is a similarity measure.
Many different similarity measures such as the ones introduced in
Section~\ref{sec:distance} have been used for trajectory clustering.
For partition-based clustering, the most appropriate similarity
measure used in the application scenario can vary widely (e.g.,
vehicle \cite{Chan2018,Ferreira2013,Pelekis2009,Hung2015}, soccer
player \cite{Gudmundsson2015}, cellular \cite{Wu2015}, large vessels
\cite{Yao2017}).
Parameterization can also be an important hurdle as well as several
approaches need to optimize multiple parameters in addition to $k$ in
order to produce high quality results.
As an extension of {\kmeans}, which is NP-hard when computing an exact
solution, is partition-based trajectory clustering, which can also be
computationally intractable for large collections when certain
similarity measures are required.
So, often the only solution is to find ways to reduce the number of
trajectories compared through similarity thresholding
\cite{Pelekis2009} or through approximation ratios if an exact
algorithm, which cannot be achieved in polynomial time
\cite{Chan2018}.

One possible approach is to use {\kpath}
clustering~\cite{Wangaclustering}, which is an extension of {\kmeans}
for trajectories on a road network.
The key idea in {\kpath} is to use a quasilinear similarity
measure-\ebd and prune the search space based with an inverted index
and exploit the metric properties of {\ebd}.
As we described in Section~\ref{sec:distance}, {\ebd} computes
similarity by performing a set intersection over edges, which
substantially improves the performance of pairwise similarity
computation, and also maintains comparable precision to common
measures that have a quadratic computational complexity.
Then, an indexing technique based on an edge inverted index and a tree structure for metric similarity measure is used to further reduce the number
of similarity computations in the assignment and refinement phases.

\subsubsection{Density-based Clustering}
Density-based trajectory clustering first finds dense ``segments'',
and then connects these segments to produce representative routes.
However, the most appropriate approach to identifying dense segments
highly dependent on properties of the data.
\traclus~\cite{Lee2007} is the most cited trajectory clustering
algorithm which solves this problem in two steps, each trajectory is
partitioned into a set of line segments (sub-trajectory), and then,
similar line segments are grouped together into clusters.
A distance threshold $\tau$ is the most common parameter used in
density-based clustering solutions, which specifies the neighborhood
radius.

Based on \traclus~\cite{Lee2007}, \citet{Li2010b} proposed an
incremental clustering framework to allow new trajectories to be
added to a database, and a new parameter is further introduced for a new
step-\textit{generating micro-clusters before the final clusters}.
Similarly, \citet{Agarwal2018} reduced the clustering problem
to finding a frequent path in a graph, which allowed standard
graph traversal techniques to be applied to identify
\textit{pathlets} that could be used to represent each cluster with common sub-trajectories.
{Each cluster's pathlet is defined as a sequence of points that is not necessarily a subsequence of an input trajectory.
	The main difference with \traclus is that the pathlets are the result of optimizing a single objective function to best represent the trajectories, while \traclus generates a set of common segments through density-based clustering, then connects them to form several common paths to represent the trajectories.}
Unfortunately, finding pathlets has been proved to be an NP-hard
problem, so exact solutions are intractable. 


\subsection{Trajectory Classification}
For urban data, there are mainly two types of trajectory
classification problems: 1) similarity-based classification
\cite{Lee2008}; 2) transportation modes and
activities classification \cite{Zheng2008,Dabiri2019} which was
reviewed recently by \cite{Zheng2015b}.
Next, we will mainly focus on the similarity-based
classification, and also introduce a new classification task for
travel-based inferences~\cite{Nair2019,Chen2019b,Gong2016,Wang2017f}.

\subsubsection{General Similarity-based Classification}
Several studies have investigated trajectory classification problems which
also require extensive similarity computations
\cite{Lee2008,Patel2012, Sharma2010}.
The main distinction in trajectory clustering is that classification
assigns a label to an individual trajectory based on its features,
while clustering is conducted over all items in a dataset.
Specifically, {\citet{Lee2008}} proposed a feature generation
framework for trajectory classification, where two types of
clustering, region-based and trajectory-based \cite{Lee2007}, are
used to generate features for traditional classifiers such as
decision tree or an SVM.
\citet{Sharma2010} proposed a nearest neighbor classification for
trajectory data by computing distance with trajectory sampled and
then choosing the nearest trajectory as the label.
\citet{AndresFerrero2018} found relevant sub-trajectories
as features for robust classification, where the distance is
computed between two equal-length sub-trajectories.

\subsubsection{Trip Purpose Classification}
Inferring the purpose of a trip has potential applications for 
improving urban planning and governance~\cite{Wang2017f}, but it is
normally conducted by through manual surveys.
With the proliferation of trajectory data availability, trip purposes
can be also classified.
\citet{Gong2016} categorized the spatiotemporal characteristics of
nine daily activity types based on inference results, including their
temporal regularities, spatial dynamics, and distributions of trip
lengths and directions.
\citet{Wang2017f} proposed a probabilistic framework for inducing trip
ordering in massive taxi GPS trajectories collections.
The key idea is to augment the origin and destination with neighbor
POIs and identify POI links based on time periods.
Then the trip intents can be explained semantically.
\citet{Nair2019} learned a model to automatically infer the purpose
of a cycling trip using personal data from cyclists, GPS trajectories,
and a variety of built-in and social environment features extracted
from open datasets characterizing the streets.


}


\section{trajectory-boosted applications in urban planning}
\label{sec:application}

Large-scale trajectory data is primarily collected in urban areas,
making it a valuable tool for applications in real-time smart
cities and timely decision making.
We will focus on four categories of applications which can benefit from trajectory data.
Table~\ref{tab:applications} provides a summary of representative
work in each of these broad areas.
To intuitively illustrate the connections between each of these, we also
provide several examples in Fig.~12.
Other examples can be found in the Appendix of \cite{Wang2020}.

\hspace{-4em}\begin{table}
	\centering
	\ra{0}
	\caption{A summary of trajectory applications, where ``P2P'' denotes the point-to-point distance, and ``P2T'' denotes the point-to-trajectory distance.}
	\vspace{-1em}
	\label{tab:applications}
	\scalebox{0.75}{
		\begin{tabular}{cccccccc}
			\toprule
			\addlinespace[0pt]\rowcolor{gray!30}
			\textbf{Type} &                             \textbf{Work} & \textbf{\makecell {Map\\matching}} & \textbf{Storage} & \textbf{Measure} & \textbf{\makecell{Representative \\ query}} & \textbf{Clustering} & \textbf{Indexing} \\
			             \addlinespace[0pt]\midrule               &              \makecell{Monitoring  \\ \cite{Wang2019,Wangsheng2018,Xie2017a}} &     \cmark     &     segment      &       -        &     \makecell{Range \\query }     &      \cmark      &      \cmark       \\ 
			                       \makecell{Road \\Traffic}                        &                   \makecell{ Jam \& Flow\\ \cite{DAndrea2017,Zheng2014a,Jeung2008}} &     \cmark     &     segment      &      EBD      &     \xmark      &      \cmark      &      \cmark       \\  
			                                                      &                \makecell{Anomaly   \\\cite{Chen2012a,Wang2018f}} &     \xmark     &      point       &     EDR             &     \xmark            &                  \cmark&    \cmark               \\ \midrule
			                                                      &         \makecell{Network design  \\ \cite{Chen2014a,Liu2016,Pinelli2016}} &     \xmark     &   point               &    P2P              &   \makecell{Constrained\\ path search}              &  \cmark                &  \cmark                 \\  
			                     \makecell{Green \\Transport}                       &              \makecell{  Navigation \\\cite{Dai2016,Guo2018a}} &     \cmark     &      segment            &  \xmark                &   \makecell{Shortest \\path query}              &    \xmark              &         \cmark          \\  
			                                                      &              \makecell{ Carpooling \\ \cite{Hsieh2017,He2014}} &     \xmark     &            segment      &        P2P         & Join                &      \cmark            & \cmark                  \\ \midrule
			                                                      &          \makecell{  Trip search \\  \cite{Wang2018a,Wang2017}} &     \xmark     &        point          &    P2T             &    TkSTT             &        \cmark          &       \cmark            \\  
			                       \makecell{Tourism \\ Planning}                      &        \makecell{   Customized trip
			              \\ \cite{Wang2018b,Chen2011a}} &     \xmark     &         point         &         P2P         &        \makecell{ Frequent \\path }       &        \cmark          &             \cmark      \\ 
			                                                      &             \makecell{Interest Discovery \\\cite{Palma2008,Wang2018b}} &     \xmark     &      point            &     \makecell{ P2P \\P2T}            &           \makecell{ Multiple \\queries }        &        \cmark          &    \cmark               \\ \midrule
			                                                      &              \makecell{Billboard \\\cite{Zhang2018c,Zhang2019a}} &     \xmark     &     point             &     P2T             &     Set cover        &       \xmark           &             \cmark      \\ 
			                   \makecell{Site\\ Selection}                     &           \makecell{Charging station\\  \cite{Li2015c,Liu2017d}} &     \xmark     &     point             & \xmark                 &     \xmark            &      \xmark            &         \cmark          \\  
			                                                      &          \makecell{ Facility route\\  \cite{Bao2017a,Wang2018c,Li2018d}} &     \cmark     &  segment                &   P2T               &     \makecell{ Shortest \\path query}           &   \cmark               & \cmark                  \\ 
			                                                       \bottomrule			                                          
		\end{tabular}}
		\vspace{-1em}
\end{table}

\vspace{-1em}

\subsection{Applications in Road Traffic}
Cities and government agencies are often a valuable source of
trajectory data.
There has been an increased emphasis on collecting data from traffic
signals and other related sensors for internal auditing purposes, and
regulatory requirements often require the data collected to be made
publicly available since it was gathered using tax income.
Such data can be used for traffic monitoring \cite{Wang2019}, anomaly
detection \cite{Wang2018f}, and traffic jam and flow analysis
{\cite{Wang2013f}}.

\begin{figure*}
	\centering
\begin{tikzpicture}[scale=0.9]
\node[align=center] at (0,0) {\small Trajectory \\ \small dataset};

\draw [thick,->] (0.8,0) -- (1.2,0);

\node[align=center] at (1.8,0) {\small Map\\\small matching};

\draw [thick,->] (2.5,0) -- (3.2,1); %
\draw [thick,->] (2.5,0) -- (3.2,-1); %
\draw [thick,->] (2.5,0) -- (3.2,0); %


\node[align=center] at (4,1) {\small Input \\ \small query};
\node[align=center] at (4,0) {\small $k$};

\node[align=center] at (4,-1) {\small Incoming\\ \small trajectory};

\draw [thick,->] (4.5,0) -- (5,0); %
\draw [thick,->] (4.5,1) -- (5,1); %
\draw [thick,->] (4.5,-1) -- (5,-1); %

\node[align=center] at (6,1) {\small Similarity \\ \small computation};
\node[align=center] at (6,0) {\small Assign \\ \small trajectories};
\node[align=center] at (6,-1) {\small Coverage\\ \small check};

\draw [thick,->] (6.7,0) -- (7.2,0); %
\draw [thick,->] (6.7,1) -- (7.2,1); %
\draw [thick,->] (6.7,-1) -- (7.2,-1); %

\node[align=center] at (8,1) {\small Pruning \\ \small by index};
\node[align=center] at (8,0) {\small Refine \\ \small cluster};
\node[align=center] at (8,-1) {\small $\theta$ \\\small comparison};

\draw [thick,->] (8.7,0) -- (9.2,0); %
\draw [thick,->] (8.7,1) -- (9.2,1); %
\draw [thick,->] (8.7,-1) -- (9.2,-1); %

\node[align=center] at (10,1) {\small Related \\ \small trajectories};
\node[align=center] at (10,0) {\small $k$ paths};
\node[align=center] at (10,-1) {\small Anomaly \\ \small or not};

\draw [line width=0.3mm,-implies,double, double distance=0.3mm] (10.7,1) -- (11.1,1);
\draw [line width=0.3mm,-implies,double, double distance=0.3mm] (10.7,0) -- (11.1,0);
\draw [line width=0.3mm,-implies,double, double distance=0.3mm] (10.7,-1) -- (11.1,-1);

\node[align=center] at (12,1) {\small Monitoring \\ \small traffic \cite{Wang2019}};
\node[align=center] at (12,-1) {\small Anomaly \\ \small dection \cite{Chen2012a}};
\node[align=center] at (12,0) {\small Flow \\ \small analysis \cite{Wangaclustering}};




\end{tikzpicture}
\vspace{-0.5em}
	\caption{Pipeline decomposition of three types of work on road traffic.}
	\vspace{-2em}
\label{fig:pipeline-road}

\end{figure*}
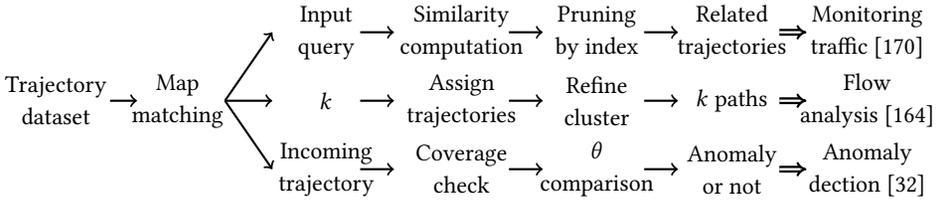

\myparagraph{An Overview}
Fig.~\ref{fig:pipeline-road} shows the pipeline of three
representative trajectory analytical tasks for road traffic.
As a common operator, mapping trajectories onto road networks not
only cleans the data, but also enables traffic jam and flow analysis
in real time.

\subsubsection{Traffic Monitoring}
{Since trajectories record the trace of vehicles in the road,
many visualization systems \cite{Chen2015} were 
developed to observe and monitor past or real-time traffic trends and movement patterns
based on trajectory data. Users can interactively explore the traffic condition in a specific area or road, and further control the traffic if necessary.}
As visualizing an entire city-wide trajectory dataset on a single screen can
be unreadable, selecting an area and identifying the trajectories
covered is a useful application of a range query, which was described
in Section~\ref{sec:spatialonly}.
In addition to a range query, a wide variety of {\em trajectory
search} queries
\cite{Chen2005a,Krogh2016,SanduPopa2011,DeAlmeida2005,Koide2015,Roh2011}
have been proposed over the years to support various applications.
To monitor all vehicles that use \textit{a Main Street}, a
\textit{{path query}} \cite{Krogh2016} can be issued.
Further, a \textit{{strict path query}} identifies
every vehicle that traverses all of \textit{a Main Street}.
\citet{Wang2019} integrated all the above queries into a real-time
traffic monitoring system, where user can interactively conduct
queries and get results in real time.

\subsubsection{Traffic Jam and Flow Analytics}

Live traffic monitoring enables analysts to track usage of
specific roads or areas in real time, and can be used to identify
traffic jams in order to notify commuters through digital traffic
signs or GPS applications.
After mapping trajectory data to a road network, {\citet{Wang2013f}}
visually inspected traffic data and verified problem areas based
on speed limit information of the streets under consideration.

Instead of inspecting individual road segments to identify
traffic jams, traffic flow analysis does not rely on domain-specific
features such as speed limits.
The goal is to discover important movement patterns of drivers such
as repeatedly using a common route composed of several different road
segments.
Several recent papers have proposed concept definitions that capture
essential properties of traffic flow analysis based on trajectories.
\citet{Gudmundsson2006} defined the concept of a \textit{flock},
which is a common sub-trajectory covered by a set of trajectories
with at least $k$ circles with the same-radius.
\citet{Jeung2008} extended the concept of a \textit{convoy}, where
the circles can have various radii.
\citet{Li2010c} proposed the concept of a \textit{swarm} to solve the
problem that a sub-trajectory has to be a continuous set of points in
the corresponding trajectory.
Rather than a single sub-trajectory, \citet{Zheng2014a} proposed the
concept of \textit{gathering}, which returns multiple
sub-trajectories.
To further avoid specifying multiple hyper-parameters, e.g., the
number of circles, {\kpath} trajectory clustering will return $k$
most representative trajectories (real paths on road network)
\cite{Wangaclustering}, which can be used in the traffic flow
analysis.



\subsubsection{Anomaly Trajectory Detection}
Anomaly or outlier trajectories in a dataset can be defined as
a trajectory that falls outside of a predefined confidence
interval w.r.t. the entire collection distribution.
It has be applied in applications such as identifying criminal
behavior in a population~\cite{Shen2015} and taxi driver fraud
detection~\cite{Wang2018f}.
A combination of similarity and clustering can be used to
solve this problem.
Computing similarity between trajectories can be used to identify
trajectories that are the most dissimilar in a dataset, and
clustering can be used to aggregate common trends that might not be
easily identifiable otherwise.
For example, given an incoming trajectory $T$ between a source $s$
and a destination $d$, anomaly detection was defined by
\citet{Chen2012a} as: 1) the anomaly score of $T$ based
on the proportion of existing trajectories in $D$ passing through $s$
and $d$ and covering $T$ completely; 2) a predefined threshold
$\theta$ to determine if $T$ is anomalous or not.
The similarity measures are based on two types of information:
1) sub-trajectories~\cite{Lee2008a}; 2) entire
trajectories~\cite{Chen2012a,Wang2018f}.
\citet{Lee2008a} detected outliers using line segments, i.e.,
a trajectory was partitioned into a set of line segments, and then,
outlying line segments were used to find trajectory outliers.
\citet{Wang2018f} utilized an edit-based similarity measure to detect
anomalies in a hierarchical cluster arrangement.

\subsection{Applications in Green Transport}
Public transportation networks such as subways, bus routes, and gas
stations provide more choices for green commuting.
Trajectory data is commonly applied in green transport (also known as ``sustainable transport''
\cite{Schiller2017}) applications to optimize network design, conduct
personalized and adaptive navigation, and carpooling.
Note that this survey focuses on algorithms and complexity,
and we also refer the interested readers to a review which covers this problem in detail
\cite{Markovic2019} for a more comprehensive view from the
perspective of road transportation agencies.

\begin{figure*}
	\centering
\begin{tikzpicture}[scale=0.9]
\node[align=center] at (0,0) {\small Trajectory \\ \small dataset};

\draw [thick,->] (0.8,0) -- (1.2,0);

\node[align=center] at (2,0) {\small Graph \\ \small construction };

\draw [thick,->] (2.5,0) -- (3.1,0); %

\node[align=center] at (4,0) {\small Edge \\\small information \\ \small enrichment};

\draw [thick,->] (0.8,0) -- (1.6,-1); %

\node[align=center] at (3.2,-1) {\small Trajectory clustering};

\draw [thick,->] (4.8,0) -- (5.3,0); %

\node[align=center] at (6,0) {\small Input OD \&\\  \small preference};

\draw [thick,->] (4.8,0) -- (5.3,1); %
\node[align=center] at (6,1) {\small Candidate \\  \small generation};

\draw [thick,->] (6.7,0) -- (7.2,0); %

\node[align=center] at (8,0) {\small Graph \\ \small traversal};

\draw [thick,->] (8.7,0) -- (9.2,0); %

\draw [thick,->] (6.7,1) -- (7.2,1); %

\node[align=center] at (8,1) {\small Route \\ \small selection};

\draw [line width=0.3mm,-implies,double, double distance=0.3mm] (8.7,1) -- (9.1,1);

\node[align=center] at (10.9,1) {\small Network planning \cite{Chen2014a}};

\draw [thick,->] (4.8,-1) -- (5.3,-1); %
\node[align=center] at (6,-1) {\small Similarity \\ \small join};

\draw [thick,->] (6.7,-1) -- (7.2,-1); %

\node[align=center] at (8,-1) {\small Rider \\ \small matching};

\draw [line width=0.3mm,-implies,double, double distance=0.3mm] (8.7,-1) -- (9.1,-1);

\node[align=center] at (10.3,-1) {\small Carpooling \cite{He2014}};

\node[align=center] at (12.3,0) {\small Navigation \cite{Dai2015}};

\node[align=center] at (10,0) {\small Optimal \\ \small route};

\draw [line width=0.3mm,-implies,double, double distance=0.3mm] (10.7,0) -- (11.1,0);





\end{tikzpicture}
\vspace{-1.5em}
	\caption{Pipeline decomposition of work on green transport.}
	\vspace{-1em}
\label{fig:pipeline-transport}
\end{figure*}
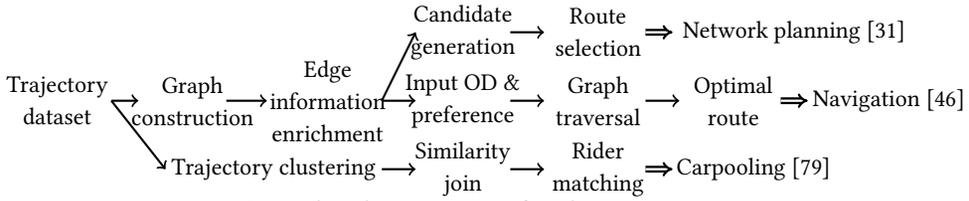

\myparagraph{An Overview} 
Based on these examples, we observe that a critical task in road networks is
trip planning or finding common routes in a graph for a passenger or
a driver, as shown in Fig.~\ref{fig:pipeline-transport}.
Complex route planning problems typically maximize a capacity or
minimize a travel time.
When viewed this way, many of the problems encountered are NP-hard,
reducible to the traveling salesmen problem, and solved using a
greedy algorithm.


\subsubsection{Transit Network Planning}
By mining taxi data, \citet{Chen2014a} approximated night
time bus route planning by first clustering all points in taxi
trajectories to determine ``hot spots'' that could be bus stops, and
then created bus routes based on the connectivity between two stops.
{\citet{Toole2015162} used census records and mobile phone location
information to estimate demand in transit routes.
Historical traffic data~\cite{6878453}, smart cards~\cite{Zhang2017},
sensors~\cite{Lv2014}, and cellular data~\cite{Pinelli2016} can
provide more comprehensive demand data and be used to further improve
data-driven transit network design.
\citet{Pinelli2016} derived frequent patterns of movement from
trajectories by computing distance and the flow gain, then merge them
to generate a network for the whole city.
A common shortcoming of these methods is that they are building a new
network, and are not applicable for most cities which already have
bus networks.
Based on human mobility patterns extracted from taxi data and smart
card data, \citet{Liu2016} proposed a localized transportation choice
model, which predicted travel demand for different bus routes by
taking into account both existing bus network and taxi routes.

\subsubsection{Adaptive Navigation}
Travel distance-based shortest path search between a source and
destination is widely used in navigation services.
It can be applied adaptively using historical trajectory data from
drivers.
With precise map matching techniques, discovering trajectory paths
in data can enrich models for a road network, e.g., dynamic time
cost of each road~\cite{Dai2016}, driver preferences
\cite{Guo2018a}. 
Using a dynamic travel time cost for each road segment, travel time
estimation \cite{Dai2016} and fastest path search \cite{Wang2019c}
is more realistic, capturing various traffic conditions, and
potentially leading to alternative shortest paths when all aspects
are considered.
\citet{Guo2018a} proposed learning to route with sparse trajectory
sets, by constructing a region graph of transfer routing preferences.
In the graph, the nodes are equal-size grids, and the edges represent
two grids crossed by more than a fixed number of trajectories.
Discovering preferences from trajectories can also be used
to support personalized route recommendation for green vehicle
routing~\cite{bektacs2016green}. 
\citet{Dai2015} developed techniques to support efficient
trajectory subset selection for following using 
driver preferences (travel distance, travel time, and consumed fuel)
and the source, destination, and departure time specified by the
driver. 
\citet{Delling2016} proposed a framework to generate personalized
driving directions from trajectories, where the path preferences 
were composed of several features, such as avoiding tolls, U-turns,
and highway versus surface streets.

\subsubsection{Carpooling}
There are two common approaches to ridesharing problems which rely
heavily on trajectory data.
The first one is grouping passengers and drivers based on their
historical trajectory data.
The underlying goal is to keep all seats in all vehicles occupied,
and is essentially a type of bin packing problem.
For example, \citet{Hsieh2017} devised a carpooling system to match
passengers and drivers based on their travel trajectories, and the
similarity measure was defined as the sum of the distance from the
pickup and drop off points to the nearest point to a driver
trajectory.
The other approach is to detect frequent routes using only rider
trajectories.
\citet{He2014} proposed a carpooling system that generates an
efficient route for dynamic ridesharing by mining the frequent
routes taken by participating riders.
\citet{Hong2017} performed rider matching by clustering trajectories,
and generated a simulation model of ridesharing behaviors to
illustrate the potential impact.
Similar to other road traffic applications, similarity computations
are performed between passenger and driver trajectories, clustering
is then used to detect the frequent routes shared by multiple
passengers in the trajectory-driven carpooling.

\subsection{Applications in Tourism Planning}
The tourism market is an enormously valuable commodity in many
countries.
For example, this market in Britain is expected to be worth more than 
£257 billion by 2025, and represents 11\% of the total GDP of the
UK.\footnote{\url{https://www.visitbritain.org/visitor-economy-facts}}
From a consumer perspective, historical trip data can also improve
overall satisfaction, and provide tools to discover more personalized
opportunities that align with individual preferences.

\subsubsection{Customizing Tours}
Searching for similar trips related to a set of given points with
keywords \cite{Wang2017,Zheng2013c,Zheng2015}, or a range
\cite{HanWZZL15} can enable more effective zero knowledge querying
capabilities, as we described in Section~\ref{sec:spatialtextual}.
In addition to spatial proximity and text relevance, other factors
can be considered, such as photos \cite{Lu2010a}.
\citet{Lu2010a} leveraged existing travel clues recovered
from 20 million geo-tagged photos to suggest customized travel route
plans according to user preferences.
\citet{Wei2013} developed a
framework to retrieve the top-$k$ trajectories that pass through
popular regions.
Rather deriving search results using only existing trips, generating
new trip plans without keywords or locations is also
possible. 
\citet{Chen2011a} discovered popular routes from spatial-only
trajectories by constructing a transfer network from trajectory data
in two steps.
First, hot areas are detected as nodes in the network; then edges are
added based on connectivity properties in the entire trajectory
collection.
Then, a network flow algorithm was proposed to discover the most
popular route from the transfer network.
\citet{Wang2018b} developed an interactive trip planning system
called TISP, which enabled interesting attractions to be
discovered, and used to dynamically modify recommendations using
click-based feedback from POIs displayed on the map.

\input{toursim-pipe}

\subsubsection{User Interest Discovery}
When a trajectory is enriched with text, it is possible to discover
more useful patterns by leveraging semantic information as well as
human interaction.
Interest discovery is a valuable tool to improve the effectiveness of
recommendations for restaurants, attractions, or public event to
tourists.
The three most common categories of interests discovery using
trajectory data for tourism are: 1) point of interest (POI)
\cite{Giannotti2007}; 2) region of interest (ROI)
\cite{Uddin2011,Jarv2018}; and 3) interactive discovery
\cite{Brilhante2015a,Wang2018a}.
\citet{Palma2008} discovered interesting locations from trajectories
using a spatiotemporal clustering method, based on the speed of
single trajectories. 
\citet{Uddin2011} presented a generalized ROI definition for
trajectories which is parameter independent, and an efficient index
over the segments by speed was proposed to find the ROIs without
scanning the whole dataset.
\citet{Brilhante2015a} utilized Wikipedia content, trajectories over
georeferenced Flickr photos, and human feedback to discover a
``budget-constrained sightseeing tour'' using a tourist's preferences
and available time.
\citet{Wang2018a} proposed a unified index, composed of inverted
index and grid index, to find POIs, including attractions, hotels,
and restaurants, to achieve fast performance for real-time response.


\subsubsection{Semantic Pattern Mining}

Generally, semantic pattern mining aims to mine the frequent movement
with descriptive text information, which is more comprehensive than a
spatial-only route.
\citet{Zhang2014a} discovered fine-grained sequential patterns which
satisfy spatial compactness, semantic consistency and temporal
continuity simultaneously from semantic trajectories.
The algorithm first groups all the places by category and retrieves a
set of coarse patterns from the database, then splits a coarse
pattern in a top-down manner.
Instead of two steps in \cite{Zhang2014a}, \citet{Kim2015b} presented
a latent topic-based clustering algorithm to discover semantic
patterns in the trajectories of geo-tagged text messages.
However, the above method can only work over trajectories with texts.
\citet{Choi2017} studied a regional semantic trajectory pattern
mining problem, aiming at identifying all the regional sequential
patterns in semantic trajectories.
Semantic pattern mining was not covered in previous
surveys~\cite{Parent2013a, Zheng2015}.
Trajectory pattern mining was grouped into four categories 
by \citet{Zheng2015}: 1) co-movement pattern which has been described
in the traffic flow analytics, e.g., the convoy, flock, group,
gathering; 2) trajectory clustering (check
Section~\ref{sec:clustering}); 3) sequential patterns which indicate
a certain number of moving objects traveling a common sequence of
locations in a similar time interval; 4) periodical patterns which
indicate periodic behaviors for future movement prediction.

\begin{table}
	\centering
	\caption{Recent work on trajectory-driven site selection.}
	\vspace{-1em}
	\scalebox{0.75}{
	\begin{tabular}{ccccccccc}
		\toprule	\addlinespace[0pt]\rowcolor{gray!30}\rowcolor{gray!30}
		& \textbf{Work} & \textbf{Constraint} & \cellcolor{gray!30} \textbf{\makecell{Extra Input Data\\besides Trajectory}}& \textbf{\makecell{Road\\ Network}} & \textbf{\makecell{The Objective \\Function}} & \textbf{\makecell{The Reduced \\NP Problem}} & \textbf{\makecell{ Accelerating \\Strategies~~~~}} & \textbf{Guarantee} \\ \addlinespace[0pt]\midrule
	\multirow{7}{*}{\hspace{-4em}$\left. {\ref{sec:charging}}\begin{array}{l}
		\\
		\\
		\\ \\ \\ \\ \\ 
		\end{array}\right\lbrace{}$} 	& \cite{Li2015c} & $k$ stations & \makecell{existing stations,\\\#charging points} & \cmark & \makecell{travel time,\\waiting time} & \makecell{ integer\\ programming} & LP-rounding & N/A \\
		 &\cite{Han2016} & $k$ stations & \makecell{+\\battery\\ performance} & \cmark & \makecell{+\\installing cost,\\charging cost,\\ waiting time} & - & \makecell{evolution \\algorithm} & - \\
		 &\cite{Liu2017d} & \makecell{for the \\whole city} & \makecell{ petroleum\\ stations, POIs,\\ real-estate} & \cmark & \makecell{revenue, \\queueing time} & \makecell{bilevel \\optimization} & \makecell{ alternating \\framework} & \makecell{ local\\ minima} \\ \midrule
		\multirow{12}{*}{\hspace{-4em}$\left.\ref{sec:billboard}\begin{array}{l}
		\\
		\\
		\\ \\ \\ \\ \\ \\ \\ \\ \\ \\ \\
		\end{array}\right\lbrace{}$} 	& \cite{Guo2017} & $k$ billboards & \makecell{bus \\advertisement, \\bus station \\audiences} & \xmark & influence & set-cover & \makecell{index,\\expansion,\\upper bound} & $1-1/e $ \\
		& \cite{Liu2017c} & \makecell{budget or\\ $k$ location} & \makecell{traffic volume,\\ speed} & \xmark & \makecell{coverage\\ value} & \makecell{ maximum\\ coverage} & \makecell{ inverted\\ index,\\greedy\\ heuristic} & $1-1/e $ \\
	&	\cite{Zhang2018c} & budget & \makecell{billboards' \\price} & \xmark & \makecell{one-time \\ impression } & \makecell{set cover\\ problem} & bound & $1-1/e $ \\
	&	\cite{Zhang2019a} & budget & \makecell{billboards' \\price} & \xmark & \makecell{ logistic \\influence} & \makecell{ biclique\\ detection} & \makecell{branch-\\and-bound} & $\frac{\theta}{2}(1-1/e)$ \\
	&	\cite{Wang2019a} & $k$ & \makecell{advertisement\\ topics,\\ traffic conditions,\\ mobility transition} & \cmark & \makecell{influence \\spread} & \makecell{weighted\\ maximum\\
		coverage} & \makecell{divide-\\and-conquer} & N/A \\ \midrule
	\multirow{7}{*}{\hspace{-4em}$\left.\ref{sec:facility}\begin{array}{l}
	\\
	\\
	\\ \\ \\ \\ \\ 
	\end{array}\right\lbrace{}$} 	&	 \cite{Bao2017a} & \makecell{ budget,\\construction \\cost,\\utilization} & \makecell{ bike\\ trajectories} & \cmark & \makecell{ beneficial \\score} & \makecell{ 0-1 knapsack\\ problem} & \makecell{ greedy \\network \\expansion,\\inverted \\index } & N/A \\
	&	\cite{Wang2018c} & \makecell{travel \\distance} & \makecell{ node\\ capacity } & \cmark & \makecell{ route \\capacity } & \makecell{constrained \\shortest\\ path search } & \makecell{divide-\\and-conquer} & exact \\
	&	 \cite{Li2018d} & \makecell{ source, \\destination } & \makecell{ parking \\edges } & \cmark & \makecell{on-road\\ travel time } & \makecell{ shortest \\path search} & \makecell{incremental \\expansion } & exact \\ \midrule
		\multirow{3}{*}{\hspace{-4em}$\left.\ref{sec:general}\begin{array}{l}
		\\
		\\
		\\ \\ \\ \\
		\end{array}\right\lbrace{}$} &	\cite{Wang2016g} & \makecell{$k=1$ } & \makecell{probability \\threshold } & \xmark & \makecell{cumulative\\ influence\\ probability } & - & \makecell{filter-\\and-validate} & exact \\
	&	 \cite{Li2016g} & $k$ & - & \cmark & \makecell{\#covered \\unique \\trajectories} & max-$k$-cover & \makecell{group \\pruning} & $1-1/e $ \\
	&	\cite{Mitra2017} & $k$ & \makecell{Existing\\facilities} & \cmark & \makecell{User \\inconvenience} & k-center & best-first & $1-1/e $ \\ 
	 \bottomrule
	\end{tabular}}
	\label{tab:site-select}
	\vspace{-1.5em}
\end{table}

\subsection{Applications in Site Selection}
As a core decision-making tool, trajectory-driven site selection has
been a crucial factor in increasing business profit and public
service quality.
Using collected trajectory data to estimate the influence of
selected sites for drivers or passengers can be applied to
problems such as charging
station placement~\cite{Li2015c}, billboard placement~\cite{Guo2017}, 
and facility route design~\cite{Wang2018c}.

\begin{definition}\textbf{(Constrained Site Selection)}
Given a set of trajectories $D$, a set of facilities $F$, a
constraint value $C$, and an influence model ($IM$), the aim of
constrained site selection is to find a subset $S \subset F$ to that
satisfies an objective function: $O = \argmax_{cost(S)<C} IM(S, D)$.
\end{definition}

Different objective functions can be defined based on the exact
scenario.
In Table}~\ref{tab:site-select}, we compare recent work in this area
in terms of constraints, input data, map matchability, objective
function, NP-hardness reduction, acceleration strategies, and
approximation guarantees.

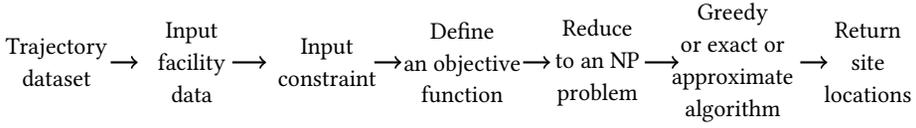
\begin{figure*}
	\centering
\begin{tikzpicture}[scale=0.9]
\tikzstyle{ann} = [draw=none,fill=none,right]

\node[align=center,rectangle] at (0,0) {\small Trajectory \\ \small dataset};

\draw [thick,->] (0.8,0) -- (1.2,0);

\node[align=center] at (2,0) {\small Input \\ \small facility \\ \small data };

\draw [thick,->] (2.6,0) -- (3.1,0); %

\node[align=center] at (4,0) {\small  Input \\ \small constraint};

\draw [thick,->] (4.7,0) -- (5.1,0); %

\node[align=center] at (6,0) {\small Define \\  \small an objective \\ \small function};

\draw [thick,->] (6.9,0) -- (7.3,0); %
\node[align=center] at (8,0) {\small Reduce  \\ \small  to an NP \\ \small problem};

\draw [thick,->] (8.7,0) -- (9.2,0); %

\node[align=center] at (10,0) {\small Greedy \\ \small or exact or \\ \small approximate \\ \small algorithm};
\draw [thick,->] (11,0) -- (11.4,0); %
\node[align=center] at (12,0) {\small Return \\\small site \\ \small locations};




\end{tikzpicture}
\vspace{-1em}
	\caption{A common pipeline decomposition of trajectory-driven site selection.}
	\vspace{-1em}
\label{fig:pipeline-site}
\end{figure*}

\myparagraph{An Overview} Fig.~\ref{fig:pipeline-site}
illustrates common connections between various site selection problems.
Site selection definitions can change based on a specific problem
scenario and may cover a singe dataset or multiple datasets.
The general problem is NP-Hard by reduction to the set cover problem,
and even approximate solutions tend to scale poorly.
Bounded or greedy algorithms can be used to accelerate the processing
along with expansion-based methods, such as estimating the bound for
all remaining set candidates by incrementally adding a single
facility and comparing it against current result to determine whether
expansion can be safely terminated.

\subsubsection{Charging Station Deployment}
\label{sec:charging}

Given the increasing popularity of electric vehicles, building more
charging stations has become a crucial problem.
Trajectory-driven charging station deployment aims to reduce the
detour distance required for charging.
\citet{Li2015c} developed an optimal charging station deployment
framework that uses historical electric taxi trajectory data,
road map data, and existing charging station information as input.
Then, it performs charging station placement and charging point
assignment (each charging station has multiple changing points).
The objective function is designed to minimize the average time to
the nearest charging station, and the average waiting time for an
available charging point, respectively, which can be reduced to the
\textit{integer linear programming (ILP)} problem
\cite{schrijver1998theory} that is NP-hard.
\citet{Liu2017d} aimed to achieve two goals: (i) the overall revenue
generated by the EVCs is maximized, subject to (ii) the overall
driver discomfort (e.g., queueing time) for EV charging is minimized.
Hence, the charging station deployment is defined as a
multiple-objective optimization problem.

\subsubsection{Billboard Placement}
\label{sec:billboard}

Billboard placement aims to find a limited number of billboards which
maximize the influence on passengers and further increase profits.
Recently, several studies
\cite{Zhang2018c,Liu2017c,Guo2017,Wang2019a,Zhang2019a} have investigated
trajectory-driven billboard placement.
\citet{Guo2017} proposed the top-$k$ trajectory influence
maximization problem, which aims to find $k$ trajectories for
deploying billboards on buses to maximize the expected influence
based on the audience.
The model of \citet{Liu2017c} was based on traffic volume.
More specifically, the algorithm counts how many
trajectories traverse an edge or vertex containing a billboard,
and then identifies the $k$ most frequent for billboard placement.
\citet{Zhang2018c} applied a model on range and one-time impressions,
constrained by total budget available for billboard placement.
\citet{Zhang2019a} proposed a logistic influence model
which solves a key shortcoming in approaches that depend on 
one-time impressions~\cite{Zhang2018c}, which did not consider the
relationship between the influence effect and the impression counts
for a single user.
\citet{Wang2019a} placed billboards in a road network, and applied a
divide-and-conquer strategy to accelerate the processing.
As we conclude in Table~\ref{tab:site-select}, billboard placement
problem definition always have a budget, and it is also general for
other site selection problem in reality as public resource allocation
also needs to be considered in the budget.

\subsubsection{Facility Route Planning}
\label{sec:facility}
Instead of pointwise candidate set selection, a facility set can also
be routes covering multiple candidates, such as planning bike lanes
and route search.
\citet{Bao2017a} designed bike lanes based on bike trajectories,
where the constraint is a budget and the number of connected
components.
A greedy network expansion algorithm was proposed to iteratively
construct new lanes to reduce the number of connected components
until the budget is met.
\citet{Wang2018c} proposed the use of MaxRkNNT for planning bus
routes between a given source and destination using a route capacity
estimation query called \rknnt.
The key constraint is travel distance, and the objective function
maximizes the estimated capacity of the routes.
Using transit trajectories, \citet{Wu2018c} defined various objective
functions for passenger preferences, and evaluated new transit routes
based on multiple factors on a real-world transport network,
including monetary cost, time cost, number of transfers, number of
choices, and transit mode.


\subsubsection{General Site Selection}
\label{sec:general}

There are several general site selection studies that are not limited
by any one scenario.
Instead of setting a cost budget, it may be more desirable to set a
parameter $k$ to choose a set of facilities from the candidate set.
Specifically, \citet{Wang2016g} aimed to identify the optimal
location ($k=1$) which can influence the maximum number of moving
objects by using the probabilistic influence.
An exact algorithm based on filtering-verification is proposed to
prune many inferior candidate locations prior to computing the 
influence.
\citet{Li2016g} identified the $k$ most influential locations
traversed by the maximum number of unique trajectories in a given
spatial region with an efficient algorithm to find the location set
using a greedy heuristic, in cases where $k$ and the spatial areas
are large.
\citet{Mitra2017} solved the trajectory-aware
inconvenience-minimizing placement of $k$ facility services which was 
proved to be NP-hard \cite{Chan2018}. 
The inconvenience is defined as the extra distance traveled to access the nearest facility location.


\section{Emerging trends on Deep Trajectory Learning}
\label{sec:deep}

Deep learning has been successfully used for trajectory data
analytics and applications for the last several years, which is attracting increasing
interest.
We group these methods into several categories, and discuss the
associated research problems in detail in this section.
%
We then cover the emerging trends in deep learning for trajectory
data.

\subsection{Trajectory Pre-processing}

\noindent {\bf Trajectory Data Generation.}
%
Generating synthetic yet realistic location trajectories plays a
fundamental role in privacy-aware analysis on trajectory data.
A generative model was proposed for generating trajectories
~\cite{Ouyang2018} based on a set of training trajectories.
The proposed solution uses generative adversarial networks (GAN) to
produce data points in a metric space, which are then transformed to a
sequential trajectory.
Experimental results show that the model is able to capture the
correlation between visited locations in a trajectory and learn the
common semantic/geographic mobility patterns from the training
trajectories.
The idea of using a GAN to generate synthetic trajectories that can
preserve the summary properties of real trajectory data for data
privacy was introduced by~\citet{Liu2018}.




\myparagraph{Trajectory Data Recovery}
As many trajectories are recorded at a low sampling rate, the
low-sampled trajectories cannot capture the correct routes of the
objects.
This problem has been studied in two settings previously, either with
road networks using map matching~\cite{Lou2009}, or without road
networks~\cite{Su2013}, as discussed in Section~\ref{sec:preprocess}.
Wang et al.~\cite{Wang2019i} aimed to recover the trajectory between
two consecutive sampled points of ``low-sampled'' trajectory data.
Using a seq2seq model, the proposed method uses spatial and
temporal attention to capture spatiotemporal correlations and
integrates a calibration component using a Kalman filter (KF) in
order to reduce the uncertainty.

\subsection{Trajectory Representation and Similarity Search }
Representation learning of trajectories aims to represent trajectories
as vectors of a fixed dimension.
Then the similarity of two trajectories can be computed based on the
Euclidean distance of their vector representations, which reduces the
complexity of similarity computation from $O(n^2)$ to $O(n)$, where
$n$ is the trajectory length.
\citet{Li2018a} proposed a seq2seq-based model to learn trajectory
representations, where the spatial proximity is taken into account
in the design of the loss function.
The trajectory similarity based on the learned representations is
robust to non-uniform, low sampling rates, and noisy sample points.
\citet{Yao2019} proposed the use of a deep metric learning framework
to approximate any existing trajectory measure, and is capable of 
computing similarity for a given trajectory pair in linear time.
The basic idea is to sample a number of seed trajectories from the
given database and then use their pairwise similarities as guide
to approximate the similarity function with a neural metric learning
framework.
The proposed solution adopts a new spatial attention memory module
that augments existing RNN for trajectory encoding.
\citet{Fu2020} proposed to exploit the road networks to learn the trajectory representation based on an encoder decoder model.
Deep trajectory representation was also extended to multi-trajectory
scenarios, such as trajectories encountered in sporting events with
several football players~\cite{Wang2019b}.
\citet{Wang2020c} applied deep reinforcement learning to enable
sub-trajectory similarity search, by splitting every trajectory into
a set of sub-trajectories that can be candidate solutions for a query
trajectory.
By learning an optimal splitting policy, the sub-trajectory
similarity search is more efficient than heuristics-based methods.


\subsection{ Deep Trajectory Clustering, Classification, and Anomaly Detection}
\citet{Yao2017} transformed trajectories into feature sequences that
capture object movements, and then applied an autoencoder framework to
learn fixed length deep representations of trajectories for
clustering.
\citet{Xuan2016} proposed a model named DeepTransport to predict the
transportation mode such as walk, taking trains, taking buses, etc.,
from a set of individual peoples GPS trajectories. LSTM
is used to constructed DeepTransport to predict a user’s
transportation mode.
\citet{Endo2016} adopted stacked denoising autoencoder (SDA) to
automatically extract features for the transportation mode
classification problem.
\citet{Gao2017} considered a different trajectory classification
problem, namely Trajectory-User Linking (TUL), which aims to link
trajectories to users who generate them in the location-based social
networks.
An RNN based semi-supervised model was proposed to address the TUL
problem.
\citet{Liu2020} proposed a deep generative model, namely a Gaussian
Mixture Variational Sequence AutoEncoder (GM-VSAE), for online
anomalous trajectory detection based on Variational autoencoding.

\subsection{Trajectory Prediction for Human Mobility Analytics}
Recurrent neural networks (RNN) are widely used for trajectory
prediction, and most of the work reviewed also learns user
representations from trajectories to capture user preferences in addition
to the spatio-temporal contexts.
\citet{Liu2016c} proposed Spatial-Temporal RNN (ST-RNN) for the next
location prediction of a given user using an RNN.
An ST-RNN model local temporal contexts, periodical temporal contexts,
and geographical contexts to learn the representation of users under
specific contexts.
%
\citet{Zhou2018a} adopted a seq2seq encoder-decoder framework
consisting of two encoders and two decoders to predict future
trajectories.
This method does not explicitly learn user representations.
%
\citet{Wu2017a} designed two RNN models to model trajectories
that consider road network constraints on trajectories, and can be
used to predict the destination of a trajectory.

DeepMove~\cite{Jin2018} is based on a multi-modal embedding RNN to
capture the complex sequential transitions by jointly embedding
multiple factors such as time, user, and location.
DeepMove also applies
attention mechanism to capture the periodical effect of mobility.
%
\citet{Chen2020} proposed a context-aware deep model called DeepJMT
for jointly performing mobility prediction (to know where) and time
prediction (to know when).
DeepJMT captures the user’s mobility regularities and temporal patterns using RNN, captures spatial context, periodicity context and social and temporal context using various mechanisms, e.g., co-attention mechanism to capture, and make time prediction using temporal point process.

A convolutional neural network (CNN) can also be used for trajectory
prediction.
\citet{Pozzi2018} proposed a CNN-based approach for representing
semantic trajectories and predicting future locations.
The semantic trajectories are represented as a matrix of semantic
meanings and trajectory IDs.
A CNN is applied to the matrix to learn the latent features for next
visited semantic location prediction.
\citet{Gao2019c} developed a deep generative model called Variational
Attention based Next (VANext) POI prediction, which simultaneously
learns implicit relationships between users and POIs, and captures
sequential patterns of user check-in behavior for next POI
prediction.
A CNN is used to capture long term and structural dependency among
user check-ins, achieving comparable learning ability with the
state-of-the-art RNN based methods, while significantly improving the
learning efficiency.
\citet{Lv2019} proposed to model trajectories as two-dimensional
images, and employed CNN for trajectory destination prediction.

\subsection{ Travel Time and Route Estimation}
\citet{Wang2018} aimed to estimate the travel time of a path from the
mobility trajectory data.
Their approach used a CNN and an RNN to capture the features of a
path, and employed a multi-task learning component to estimate the
travel time.
\citet{Zhang2018h} developed a bidirectional LSTM based deep model,
called {DeepTravel}, to estimate the travel time of a path from the
historical trajectories.
\citet{Li2019c} proposed a deep generative model, DeepGTT, to learn
the travel time distribution for a route by conditioning on the
real-time traffic, which is captured by the trajectory data.
\citet{Lix2020} proposed a deep probabilistic model, DeepST, which
unifies three key explanatory factors, the past traveled
trajectories, the impact of destination, and real-time traffic for
the route decision of a pair of source and destination.
{\citet{Yuan2020} estimated the travel time of an origin-destination
pair at a certain departure time, and proposed a neural network based
prediction model.
This model exploits the fact that for a past OD trip its travel
time is usually affiliated with the trajectory it travels along,
whereas it does not exist during prediction.}



\section{Future Research Directions and Open Issues}
\label{sec:future}

Trajectory data has inspired many important research problems and
applications in both academia and industry.
However, open-source and commercial systems that can support the
entire pipeline of trajectory data management and analytics are still
non-existent.
To better cater to future applications, a unified trajectory data
management system would be an important contribution to the
community.
Desirata of such a system include:
\begin{itemize}
\item \textbf{Data Cleaning}: Map matching, re-sampling, and
calibration are fundamental building blocks in efficient and
effective analytics.
However, current solutions need to input a road network dataset given
by users manually, and also need to input several data-dependent
parameters.
An automated data cleaning pipeline for trajectory data is highly
desirable as manual data cleaning is time-consuming and often not
reproducible.
\item \textbf{Trajectory Data Repositories}: Cleaning trajectory is
computationally expensive.
Publicly available data repositories of raw and cleaned data would
not only improve reproducibility, it would also stimulate new
research in the area.
A standard data format for trajectory data is also not currently
defined.
Even though LineString and GPX can be used, their use is limited.
Labeled datasets for similarity search and classification problems
are also non-existent, hampering progress on this important problem.

\item \textbf{Data Integration and Operator Support}: In more
advanced data analytics applications, trajectory data is often
heterogeneous.
Data integration between trajectory databases and other databases,
including spatial databases (point data) and graph databases (road
networks) would be a valuable contribution.
Trajectory data from multiple devices, including cameras, UAV, and
RFID can also be integrated to build more comprehensive profiles of a
city.
Unstructured metadata from various sources also plays a vital role
when apply trajectory data in real-world application.
This data might include speed limits, transit timetables, and public
service opening hours. 
While this information is used commonly in commercial applications,
the cost of using proprietary data sources is often too high
in the research community.

\item \textbf{Performance Benchmarks}: 
A trajectory performance benchmark similar to the TPC
benchmark\footnote{\url{http://www.tpc.org/information/benchmarks.asp}}
would greatly improve the quality of empirical comparisons of new
algorithms in research papers.
There are many different distance measures and indexing data
structures being applied to trajectory search with little
understanding of their true performance characteristics.
For efficiency, pruning power and I/O are important factors in total
running time; for the effectiveness, no ground-truth means that
search quality cannot truly be measured.

\item \textbf{Parameterization}: Parameter selection plays a critical
role in system performance of trajectory analytics.
Parameter-free algorithms are highly desirable, but not always
possible.
Automated parameter selection is an
essential requirement for automatic databases, and
for intelligent trajectory analytics.
It would also be interesting to investigate how to automate bounding
approximation ratios can be achieved during algorithm design.

\item \textbf{Deep Trajectory Learning}: Deep learning has made some progress in synthetic data generation, 
	representation learning, and mobility prediction. More advanced tasks, 
	including query optimization and learned index, can be investigated to 
	improve the query performance. Another promising topic can be the 
	online deep trajectory learning to meet the demand for timely decision 
	making over dynamic trajectory data.

\item \textbf{Self-Driving Trajectory}: {Self-driving \cite{Ma2018,Chang2019} will be one of the primary sources of trajectories in the near future.
Different from the typical long-term trajectories being currently analyzed in road-network which is the focus in this survey, self-driving trajectories will be short-term lane-level traces with high sampling rates, e.g., each trajectory only lasts for five seconds \cite{Chang2019}. Due to the requirement of real-time response for safe driving, managing and analyzing such trajectories will be crucial and challenging, especially in a streaming scenario.}

\item \textbf{Trajectory Analysis for Public Health}: Human movement data recorded by mobile phones has been analyzed to control Malaria in Africa \cite{Wesolowski2012} and dengue epidemics in Pakistan \cite{Wesolowski2015}. With mass outbreaks of diseases such as COVID-19 \cite{covid19,covid19-2} in dense urban areas, collecting and analyzing the trajectories of infected people and building real-time warning mechanisms will play an important role in disease traceability, disease control, and emergency response.
\end{itemize}

\section{Conclusion}
\label{sec:conclude}
In this survey, we have reviewed recent progress in trajectory data
management and learning.
We first presented an overview of trajectory data management and
urban applications.
Then, we categorized the problems based on components and operators
shared across the tasks.
Similarity measures, top-$k$ similarity search, and fast clustering
are all widely used for many different problems and scenarios, and
were the focus of this study.
Finally, we reviewed important research advances on four common
analytics applications of trajectories for real-time smart cities.
New applications emerge daily that leverage trajectory data, such as
deep trajectory learning, and we hope this survey can provide readers
an overview of the landscape of trajectory data management and
applications.
Perspectives on how to choose the most appropriate solution for
pre-processing, storage, search, and advanced analytics to achieve
high efficiency and effectiveness can be applied in many different
problem domains beyond road networks.


\vspace{-0.5em}
 \begin{acks}
Zhifeng Bao is supported in part by ARC DP200102611, DP180102050, and a Google Faculty Award.
J. Shane Culpepper is supported in part by ARC DP190101113.
Gao Cong is supported in part by a MOE Tier-2 grant MOE2019-T2-2-181, and a MOE Tier-1 grant RG114/19.
\end{acks}
\bibliographystyle{ACM-Reference-Format}
\bibliography{library,url}

\newpage

\appendix
\section{Supplemental figures}
\label{sec:app}

\begin{figure}[h]
	\begin{minipage}{0.38\textwidth}
		\centering
				\vspace{1em}
		\includegraphics[height=3.3cm]{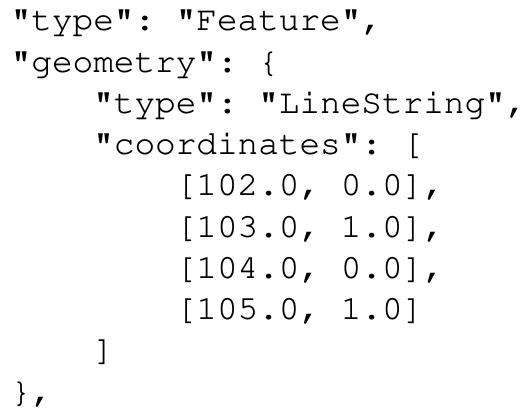}
			\vspace{1em}
		\caption{An example of \textit{LineString} with four points in the GeoJSON format.}
		\vspace{-0.5em}
		\label{fig:linstring}
	\end{minipage}\hfill
	\begin{minipage}{0.58\textwidth}
		\centering
		\includegraphics[height=3.9cm]{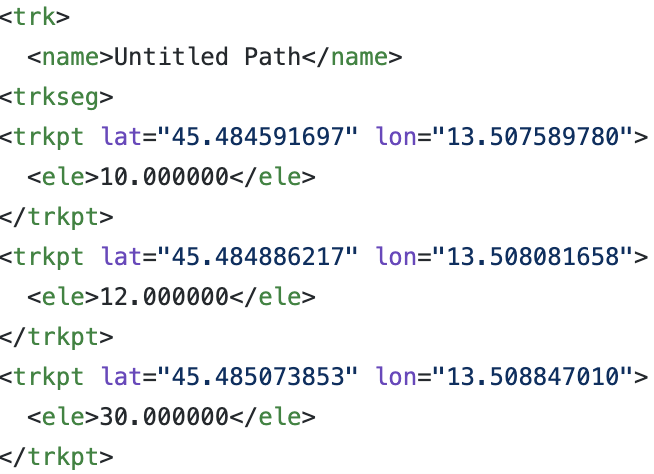}
		\vspace{0.5em}
		\caption{An example of \textit{GPX} format \cite{foster2004gpx} widely used in OpenStreetMap Public GPS trace \cite{osm}.}
		\vspace{-0.2em}
		\label{fig:gpx}
	\end{minipage}
\end{figure}

\begin{figure}[h]
	\centering
	\subfigure[Road traffic monitoring with range queries on the Porto collection \cite{Wang2019} ]{\label{fig:a}\includegraphics[width=0.39\textwidth]{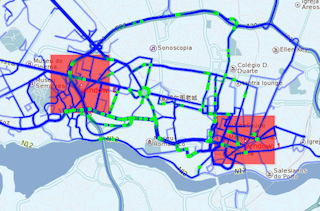}}\hspace{2em}
	\textbf{\subfigure[Transit network planning with commuter trajectory data to cater to new travel demands \cite{Wang2018c}]{\label{fig:b}\includegraphics[width=0.55\textwidth]{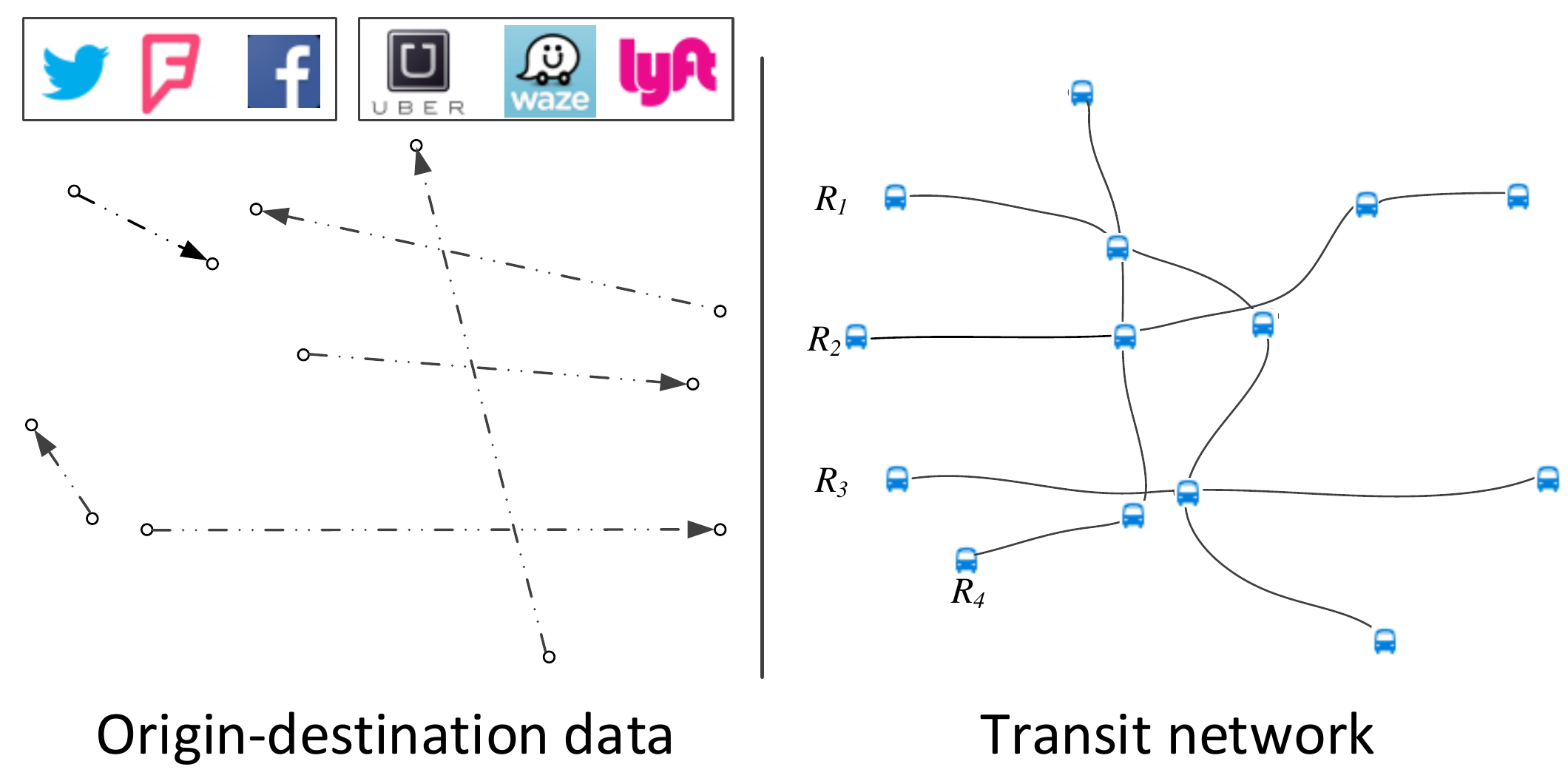}}}\\ \vspace{2em}
	\subfigure[Trajectory-based queries for travel planning and interest discovery for a tourist visiting Los Angeles~\cite{Wang2018a}]{\label{fig:c}\includegraphics[width=0.61\textwidth]{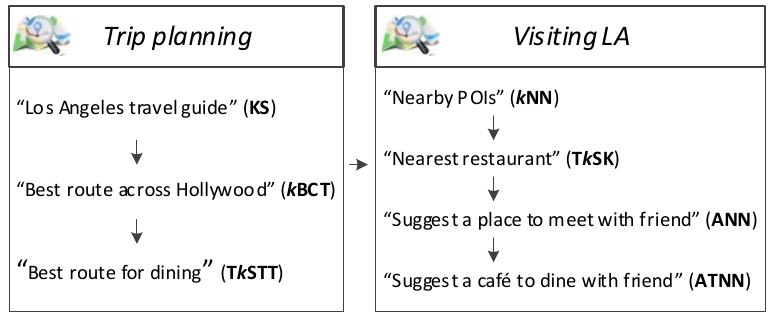}}\hspace{2em}
	\subfigure[Site selection of billboards to maximize commuter views~\cite{Zhang2019a}]{\label{fig:d}\includegraphics[width=0.32\textwidth]{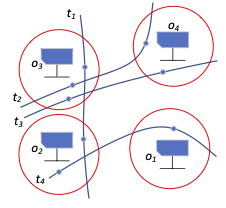}}
	\caption{Examples of four urban applications for trajectory data.}
	\label{fig:four-apply}
\end{figure}
\newpage

\end{document}